\pdfoutput=1
\documentclass[prb,twocolumn,showpacs,amsmath,amssymb,floatfix]{revtex4}

\usepackage{graphicx}%Include figure files
\usepackage{dcolumn}%Align table columns on decimal point
\usepackage{bm}% bold math
\usepackage{amssymb}

\begin{document}

\title{Signatures of electron-magnon interaction in charge and spin currents through magnetic tunnel junctions: A nonequilibrium many-body perturbation theory approach}

\author{Farzad Mahfouzi}
\email{mahfouzi@udel.edu}
\affiliation{Department of Physics and Astronomy, University of Delaware, Newark, DE 19716-2570, USA}
\author{Branislav K. Nikoli\' c}
\email{bnikolic@udel.edu}
\affiliation{Department of Physics and Astronomy, University of Delaware, Newark, DE 19716-2570, USA}

\begin{abstract}
We develop a numerically exact scheme for resumming certain classes of Feynman diagrams in the self-consistent perturbative expansion for the electron and magnon self-energies in the nonequilibrium Green function  formalism applied to a coupled electron-magnon (\mbox{e-m}) system driven out of equilibrium by the applied finite bias voltage. Our scheme operates with the electronic and magnonic GFs and the corresponding self-energies viewed as matrices in the Keldysh space, rather than conventionally extracting their retarded and lesser components, which greatly simplifies translation of diagrams into compact mathematical expressions and their computational implementation. This is employed to understand the effect of inelastic \mbox{e-m} scattering on charge and spin current vs. bias voltage $V_b$ in F/I/F (F-ferromagnet; I-insulating barrier) magnetic tunnel junctions (MTJs), which are modeled on a quasi-one-dimensional (quasi-1D) tight-binding lattice for the electronic subsystem and quasi-1D Heisenberg model for the magnonic subsystem. For this purpose, we evaluate Fock diagram for the electronic self-energy and the electron-hole polarization bubble diagram for the magnonic self-energy. The respective electronic and magnonic GF lines within these diagrams are the fully interacting ones, thereby requiring to solve the ensuing coupled system of nonlinear integral equations self-consistently. Despite using the quasi-1D model and treating \mbox{e-m} interaction in many-body fashion only within a small active region consisting of few lattice sites around the F/I interface, our analysis captures essential features of the so-called zero-bias anomaly observed [Phys. Rev. B {\bf 77}, 014440 (2008)] in both MgO- and AlO$_x$-based realistic 3D MTJs where the second derivative $d^2 I/dV_b^2$ (i.e., inelastic electron tunneling spectrum) of charge current exhibits sharp peaks of opposite sign on either side $V_b =0$. We show that this is closely related to substantially modified magnonic density of states (DOS) after \mbox{e-m} interaction is turned on---the magnonic bandwidth over which DOS is non-zero becomes broadened, thereby making \mbox{e-m} scattering at arbitrary small bias voltage possible, while DOS also acquires peaks (on the top of a continuous background) signifying the formation of quasibound states of {\em magnons dressed by the cloud of electron-hole pair excitations}. We also demonstrate that the sum of electronic spin currents in all of the semi-infinite leads attached to the active region quantifies the loss of spin angular momentum carried away from the active region by the magnonic spin current.
\end{abstract}

\pacs{72.25.-b, 72.10.Di, 72.10.Bg}
\maketitle

\section{Introduction}\label{sec:intro}

Magnetic tunnel junctions (MTJ) are layered heterostructures in which an insulating tunnel barrier (I) separates two ferromagnetic
layers (F). They have been the subject of vigorous research in both fundamental and applied physics since they exhibit effects
like tunneling magneto-resistance (TMR)~\cite{Tsymbal2003} and spin-transfer torque (STT),~\cite{Ralph2008,Brataas2012} as well as quantum size effects
in electron transport (even at room temperature) when normal metal (N) layer is inserted.~\cite{Yuasa2002} From the fundamental viewpoint, these effects represent examples of nonequilibrium quantum many-body systems with an interplay of fast conduction electrons carrying spin current and slow collective magnetization,
while from the viewpoint of applications they play an essential role in developing magnetic sensors, random access memory, novel programmable logic devices,
resonant-tunneling spin transistors and nanoscale microwave oscillators with ultrawide operating frequency ranges.~\cite{Katine2008}

\begin{figure}
\includegraphics[scale=0.5,angle=0]{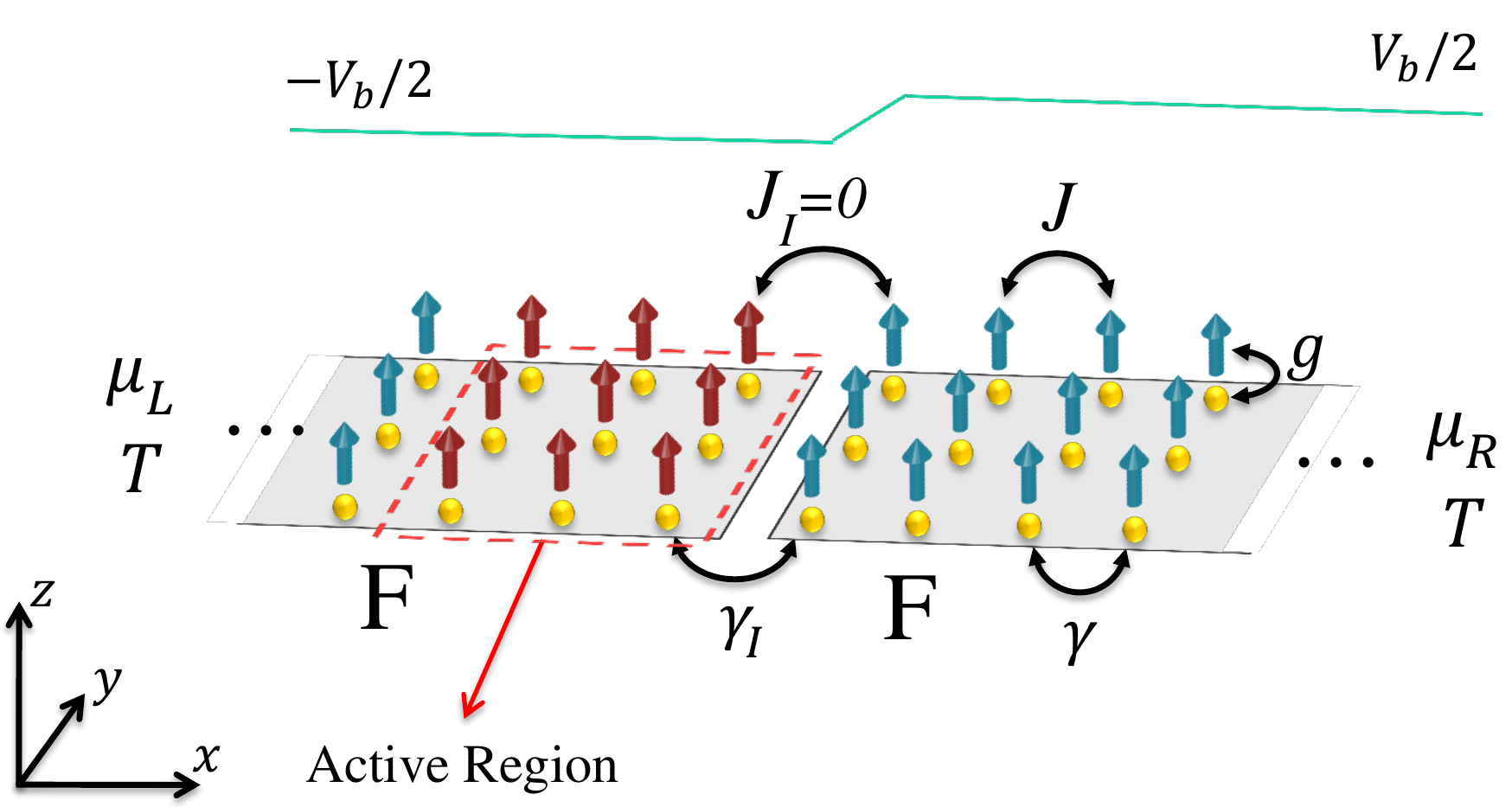}
\caption{(Color online) Schematic view of a quasi-1D model of F/I/F MTJ where the left semi-infinite ideal F lead, modeled as spin-split tight-binding lattice of size $\infty \times N_y$, is attached via hopping $\gamma$ to an active region consisting of $N_x \times N_y$ lattice sites (we use $N_x=3$ and $N_y=1$ or $N_y=3$ in the calculations below). The right semi-infinite lead is attached to the active region via smaller hopping $\gamma_I=0.1\gamma$ that simulates the tunnel barrier $I$. The same sites also host localized spins which are coupled to each other via the ferromagnetic coupling $J>0$ in the Heisenberg model. The local coupling between the spin of conduction electrons and localized spin on each site is of strength $g$. The left and right semi-infinite leads are assumed to terminate into macroscopic Fermi liquid reservoirs held at electrochemical potentials $\mu_L$ and $\mu_R$, respectively, whose difference sets the bias voltage $eV_b = \mu_L - \mu_R$. The voltage profile across MTJ is shown on the top. The left F layer is assumed to be attached at infinity to a macroscopic reservoir of magnons held at temperature $T$.}
\label{fig:fig1}
\end{figure}

The STT is a phenomenon in which a spin current of sufficiently large density injected into F layer either switches its magnetization from one static configuration to another or generates a dynamical situation with steady-state precessing magnetization.\cite{Ralph2008,Brataas2012} The origin of STT is the absorption of the itinerant flow of angular momentum components normal to the magnetization direction. For F$^\prime$/I/F MTJs illustrated in Fig.~\ref{fig:fig1}, where the reference F$^\prime$ layer with fixed magnetization ${\bf m}^\prime$ plays the role of an external spin-polarizer and right F layer has free magnetization ${\bf m}$, it is customary to analyze the parallel \mbox{${\bf T}_{\parallel}=a_J {\bf m} \times ({\bf m} \times {\bf m}^\prime)$} and perpendicular \mbox{${\bf T}_{\perp}=b_J {\bf m} \times {\bf m}^\prime$} components of the STT vector ${\bf T} = {\bf T}_{\parallel} + {\bf T}_{\perp}$. These two terms act differently on magnetization dynamics---$a_J$ effectively changes the magnetic damping (i.e., an antidamping or additional damping depending on the current polarity), whereas $b_J$ acts like a magnetic field.~\cite{Ralph2008}

A majority of theoretical studies of TMR or STT effects has assumed phase-coherent tunneling of non-interacting quasiparticles. For example, such approaches~\cite{Butler2001,Mathon2001} have led to a remarkable prediction of very large TMR ratio $\simeq 4000$\% at zero bias voltage
for clean epitaxial MgO-based MTJs. The TMR ratio is defined by $\mathrm{TMR}=(R_\mathrm{AP}-R_\mathrm{P})/R_\mathrm{P}$, where $R_\mathrm{P}$
is the resistance for parallel orientations of two magnetizations in F/I/F MTJ and $R_\mathrm{AP}$ is the resistance when they are antiparallel. These
predictions have ignited large experimental efforts that have eventually reached TMR ratios of more than 1000\% at low temperatures and $\simeq 600$\% at room temperature for well-oriented MgO barriers with stress relaxation.~\cite{Ikeda2008}  The phase-coherent calculations---such as numerical ones based on the nonequilibrium Green function (NEGF) formalism~\cite{Stefanucci2013} combined with simplistic tight-binding Hamiltonians~\cite{Theodonis2006,Tang2009} or first-principles obtained Hamiltonians;~\cite{Heiliger2008} as well as analytical ones~\cite{Xiao2008a} based on the scattering approach---have been able to capture the dependence of $a_J$ and $b_J$ on the bias voltages \mbox{$V_b \lesssim 0.2$ V} in MgO-based MTJs.~\cite{Kubota2008} However, such theories~\cite{Theodonis2006,Tang2009,Xiao2008a} including {\em only} elastic electron tunneling start to deviate from experimental findings at higher bias voltages, which is particularly pronounced~\cite{Li2008b,Park2011a,Wang2011} for $b_J$ (playing a significant role during magnetization switching at \mbox{$V_b \simeq 1.0$ V}).

The inelastic electron-magnon (\mbox{e-m}) or electron-phonon (\mbox{e-ph}) scattering could account for these discrepancies.~\cite{Li2008b,Park2011a} In particular, since magnon bandwidth is usually of the order of $\simeq 100$ meV, at high bias voltages multiple magnon scattering events can be excited.~\cite{Balashov2008} Also, energy dependence of the magnon density of states (DOS) probed~\cite{Balashov2008} at finite bias voltage is intimately linked to the evolution of the magnetization during current-driven switching when going beyond the macrospin approximation.~\cite{Strachan2008,Brataas2006a}

In fact, even at small bias voltage thermally excited magnons affect TMR (e.g., emission or absorption of magnon at F/I interface reduces the effective
spin polarization of electrons incoming from F leads in Fig.~\ref{fig:fig1}), so that TMR decreases with increasing temperature.~\cite{Drewello2008,Khan2010} Although thermally induced change of the resistance is different for AlO$_x$- and MgO-based MTJs, their inelastic tunneling spectra~\cite{Reed2008} (IETS) shows very similar properties. That is, plotting the second derivative $d^2I/dV_b^2$ of current vs. bias voltage in MTJs  reveals zero bias anomaly (ZBA) where peaks (see, e.g., Fig. 2 in Ref.~\onlinecite{Drewello2008}) of opposite sign appear at $V_b \simeq \pm 10$ mV and are related to magnons. Also, additional phonon peaks are found~\cite{Drewello2008} at \mbox{$V_b \simeq \pm 81$ mV} for the MgO-based MTJs or at \mbox{$V_b \simeq \pm 120$ mV} for AlO$_x$-based MTJs.

Theoretical efforts to capture \mbox{e-m} inelastic scattering effects on TMR, ZBA and STT have thus far utilized simplified frameworks~\cite{Levy2006,Manchon2009a} which {\em cannot} deal with multiple scattering events, backaction of magnons driven far from equilibrium and energy dependence of magnonic DOS. Such effects can be taken systematically and rigorously into account at arbitrary temperature or bias voltage by using the NEGF formalism coupled with perturbation expansion of electron or magnon self-energies in the presence of their mutual interaction in terms of the respective Feynman diagrams.~\cite{Stefanucci2013} In equilibrium problems, like that of magnetic polaron, electronic self-energy has been constructed by considering large set of diagrams involving an arbitrary number of \mbox{e-m} scattering vertices between the emission and absorption vertices.~\cite{Richmond1970} However, using the same diagrams within the NEGF framework would violate charge conservation, yielding different charge currents in the left and right lead of a two-terminal device at finite bias voltage.

One of the conserving approximations is the so-called self-consistent Born approximation (SCBA) where one considers Hartree and Fock diagrams for the electronic  self-energy which corresponds to perturbation in the order $\mathcal{O}(g^2)$ where $g$ is the strength of \mbox{e-m} interaction. Evaluation of these diagrams for a systems defined on the lattice hosting orbitals in real space is computationally very demanding due to the fact that GF lines in the diagrams of nonequilibrium many-body perturbation  theory (MBPT) are fully interacting (or dressed), so that self-energy matrix becomes a functional of the GF matrix. This generates a coupled system of nonlinear integral equations which has to be solved by performing multiple integrations over each matrix element until the self-consistency is achieved. Such route has been undertaken in only a handful of studies where further simplifications (such as using dispersionless magnons, $\omega_\mathbf{k}=\omega_0$) were utilized.~\cite{Reininghaus2006}

Here we discuss in Sec.~\ref{sec:diagrammatics} how to construct the electronic self-energy and GF within SCBA, together with the magnonic self-energy within the electron-hole (e-h) polarization bubble approximation which takes into account influence of electrons on magnons while inserting the dressed magnonic GF into SCBA diagrams. Thus, consideration of such diagrams is akin to the self-consistent GW treatment of the one-particle electronic self-energy due to electron-electron interaction out of equilibrium.~\cite{Thygesen2008} Our approach treats these quantities as matrices~\cite{Kita2010} in the Keldysh space, rather than following the commonly used route based on Langreth rules to manipulate expressions involving products of their submatrices.~\cite{Stefanucci2013} This formalism is then applied to a many-body Hamiltonian, introduced in Sec.~\ref{sec:hamiltonian}, of an interacting \mbox{e-m} system defined on the real-space lattice describing MTJ that is brought out of equilibrium by the applied finite bias voltage. Despite using a quasi-one-dimensional (quasi-1D) model for MTJ illustrated in Fig.~\ref{fig:fig1}, where \mbox{e-m} interaction is treated diagrammatically only within few lattice sites (denoted as ``active region'' in Fig.~\ref{fig:fig1})  of the left F layers, charge current versus bias voltage and its second derivative obtained in Sec.~\ref{sec:zba} capture essential features of ZBA observed in realistic junctions. The sum of spin currents carried by electrons in the non-interacting F leads attached to this active region is non-zero which, therefore, allows us to quantify in Sec.~\ref{sec:zba} the amount of lost angular momentum of electronic subsystem that is carried away by magnonic spin current. We conclude in Sec.~\ref{sec:conclusions}.

We also provide two Appendices. Appendix~\ref{sec:conservation} proves that charge current is conserved when electronic GF is computed by including the Fock diagram only, on the proviso that its electronic GF line is dressed (thereby requiring self-consistency) while its magnonic GF line can be bare or dressed. Appendix~\ref{sec:ap_hilbert} shows numerical implementation of the Hilbert transform, utilized to obtain results in Sec.~\ref{sec:zba}, for a function computed on the mesh of adaptively selected energy points which can be arbitrarily spaced form each other.

\section{Hamiltonian for coupled electron-magnon system within MTJ}\label{sec:hamiltonian}

To make the discussion transparent, we focus on the particular example of \mbox{e-m} interacting many-body system out of equilibrium which emerges within the quasi-1D model of a two-terminal F/I/F MTJ depicted in Fig.~\ref{fig:fig1}. This can be described by the following Hamiltonian
\begin{align}\label{eq:emhamiltonian}
\hat{H} = &  \sum_{i\sigma} \left(\varepsilon_i {\bm \sigma}_0  +\frac{\Delta}{2} m_z^i {\bm \sigma}^z_{\sigma\sigma} \right)_{\sigma\sigma} \hat{c}^{\dagger}_{i\sigma}\hat{c}_{i\sigma} \nonumber  \\
&+ \sum_{\langle ij \rangle, \sigma}(\gamma_{ij} \hat{c}^{\dagger}_{i\sigma}\hat{c}_{j\sigma}+\mathrm{H.c.})  \\
&-\frac{1}{2}\sum_{\langle ij \rangle} J_{ij} \hat{\vec{S}}_i \cdot \hat{\vec{S}}_j - \frac{1}{2} E_Z \sum_i (\hat{S}_i^z)^2 + g \sum_{j}  \hat{\vec{S}}_j \cdot \hat{\vec{s}}_j \nonumber
\end{align}
The first term in Eq.~\eqref{eq:emhamiltonian} accounts for the on-site potential due to the voltage profile shown in Fig.~\ref{fig:fig1}, as well as for the coupling of itinerant electrons to collective magnetization described by the material-dependent exchange potential $\Delta=0.75$ eV. We use the standard notation $\vec{\bm \sigma}=({\bm \sigma}^x,{\bm \sigma}^y,{\bm \sigma}^z)$ for the vector of the Pauli matrices, where ${\bm \sigma}^{x,y,z}_{\sigma \sigma'}$ are their matrix elements, and ${\bm \sigma}_0$ is the unit $2 \times 2$ matrix. Here $m_z^i=1$ or $m_z^i=-1$ depending on whether the magnetization of the left or right F layer is parallel or antiparallel to the $z$-axis, respectively. The second term in Eq.~\eqref{eq:emhamiltonian} describes hopping of electrons between single $s$-orbitals located on the tight-binding lattice sites, where $\hat{c}^{\dagger}_{i\sigma}$ ($\hat{c}_{i\sigma}$) creates (annihilates) electron on site $i$ in spin state \mbox{$\sigma = \uparrow,\downarrow$} and $\gamma_{ij}$ is the nearest-neighbor hopping parameter. We set \mbox{$\gamma_{ij}=\gamma=1$ eV} for all pairs of lattice sites, except for the last row of sites of the left F layer and the first row of sites of the right F layer where \mbox{$\gamma_{ij}=\gamma_I=0.1$ eV} (in the case of $N_x \times N_y \equiv 3 \times 1$ active region in Fig.~\ref{fig:fig1}) or \mbox{$\gamma_{ij}=\gamma_I=0.3$ eV} (in the case of $N_x \times N_y \equiv 3 \times 3$ active region in Fig.~\ref{fig:fig1}) simulates the presence of the tunnel barrier $I$.

The third term is the Heisenberg model~\cite{Rastelli2013} describing interaction between spin operators $\hat{\vec{S}}_i$ and $\hat{\vec{S}}_j$ localized on the nearest-neighbor sites of the same tight-binding lattice, where ferromagnetic coupling is set as \mbox{$J_{ij}=J=1 \times   10^{-3}$ eV}, except for the I region where $J_{ij}=J_I \equiv 0$. The fourth term with \mbox{$E_Z=2 \times 10^{-3}$ eV} is introduced to select energetically favorable direction (i.e., an easy-axis) for the spontaneous magnetization in the ferromagnetic layers along the $z$-axis. Finally, the fifth term describes interaction of the spin operator \mbox{$(\hat{\vec{s}}_j)_{\sigma\sigma'}=\frac{1}{2}   {\vec{\bm \sigma}}_{\sigma\sigma'} \hat{c}^{\dagger}_{j\sigma} \hat{c}_{j\sigma'}$} of conduction electrons with the localized spin operators $\hat{\vec{S}}_j$, where the coupling constant is set as $g=0.045$ eV.

The active region of MTJ in Fig.~\ref{fig:fig1}, within which NEGFs and self-energies due to \mbox{e-m} interaction are computed, consists of $N_x \times N_y$ sites enclosed in Fig.~\ref{fig:fig1}. The rest of the tight-binding sites belong to the left and right semi-infinite leads (taken into account through lead self-energies discussed in Sec.~\ref{sec:diagrammatics}). The leads are assumed to terminate at infinity into macroscopic Fermi liquid reservoirs held at electrochemical potentials $\mu_L=E_F+eV_L$ and $\mu_R=E_F+eV_R$ (the Fermi energy is chosen as \mbox{$E_F=0.5$ eV} for $N_x \times N_y \equiv 3 \times 1$ active region and \mbox{$E_F=2.65$ eV} for $N_x \times N_y \equiv 3 \times 3$ active region), whose difference sets the bias voltage $eV_b = \mu_L - \mu_R$. Concurrently, the left F layer is assumed to be attached at infinity to a macroscopic reservoir of magnons held at temperature $T$.

Using the approximate version of the Holstein-Primakoff transformations~\cite{Rastelli2013}
\begin{subequations}\label{eq:hp}
\begin{eqnarray}
\hat{S}_i^{+} & \approx & \sqrt{2S}\hat{b}_i^{\dagger},  \\
\hat{S}_i^{-} & \approx & \sqrt{2S}\hat{b}_i,  \\
\hat{S}_i^{z} & = & \hat{b}_i^{\dagger}\hat{b}_i - m_z^i S,
\end{eqnarray}
\end{subequations}
we can replace the spin operators by bosonic operators.  The approximation in Eq.~\eqref{eq:hp} is a valid when the occupation number of bosonic states at temperature $T$ is low, $\langle \hat{b}_i^{\dagger} \hat{b}_i \rangle \ll S$, where we select $S=10$. This is equivalent to saying that the left or right F layer is near its ferromagnetic ground state where $\langle \hat{S}^z_i \rangle =  S$ on all lattice sites within both the left and right F layers for the parallel (P) configuration of magnetizations in MTJ, or $\langle \hat{S}^z_i \rangle =  S$ within the left F layer and $\langle \hat{S}^z_i \rangle = - S$ within the right F layer for antiparallel (AP) configuration of magnetizations in MTJ.

This replacement allows us to rewrite the Hamiltonian in Eq.~\eqref{eq:emhamiltonian} as $\hat{H}=\hat{H}_\mathrm{e} + \hat{H}_\mathrm{m} + \hat{H}_\mathrm{e-m}$, where all three terms are now given in the second quantization
\begin{subequations}\label{eq:hamiltonian}
\begin{eqnarray}
\hat{H}_\mathrm{e} & = &  \sum_{i,\sigma} \left( \varepsilon_i {\bm \sigma}_0 + \frac{1}{2}\left( gS + \Delta \right) m_z^i  {\bm \sigma}^z \right)_{\sigma\sigma} \hat{c}^{\dagger}_{i\sigma} \hat{c}_{i\sigma} \nonumber \label{eq:he} \\
&& + \sum_{\langle ij \rangle, \sigma}(\gamma_{ij} \hat{c}^{\dagger}_{i\sigma} \hat{c}_{j\sigma} + \mathrm{H.c.}), \\
\hat{H}_\mathrm{m} & = & -S\sum_{\langle ij \rangle}J_{ij}\hat{b}_i^{\dagger}\hat{b}_j \nonumber \\
&& +S(E_Z + 2zJ) \sum_i \hat{b}_i^{\dagger} \hat{b}_i, \label{eq:hm} \\
\hat{H}_\mathrm{e-m} & = & \sqrt{\frac{S}{2}} \sum_{i} g(\hat{b}^{\dagger}_i \hat{c}^{\dagger}_{i\downarrow} \hat{c}_{i\uparrow}
+ \hat{b}_i\hat{c}^{\dagger}_{i\uparrow}\hat{c}_{i\downarrow}), \nonumber \\ \label{eq:hem}
&&+\frac{1}{2}\sum_{i\sigma} g {\bm \sigma}^z_{\sigma \sigma} \hat{b}^{\dagger}_i \hat{b}_i \hat{c}^{\dagger}_{i \sigma}\hat{c}_{i\sigma}.
\end{eqnarray}
\end{subequations}
Here $2z$ is the number of nearest neighbor sites. The many-body interaction is encoded by $\hat{H}_\mathrm{e-m}$ in Eq.~\eqref{eq:hem}, which is assumed to be non-zero only in the active device region in Fig.~\ref{fig:fig1}. Its first term has a clear physical interpretation---the spin of a conduction electron is flipped when magnon is absorbed or emitted. Since its second term (in the lowest order, its role is to renormalize the effective Zeeman splitting for electrons) is much smaller than the first term due to assumed $\langle \hat{b}_i^{\dagger} \hat{b}_i \rangle \ll S$, we ignore it in subsequent discussion.

In the rest of the system depicted in Fig.~\ref{fig:fig1}, electrons and magnons are assumed to behave as non-interacting quasiparticles. For example, in the case of $N_x \times N_y \equiv 3 \times 1$ active region, the left and right semi-infinite F leads are 1D chains whose electrons in equilibrium ($\varepsilon_i=0$) are described by $\hat{H}_\mathrm{e}$ in Eq.~\eqref{eq:he} generating dispersion $E_k=-2 \gamma  \cos(ka) \pm (gS+\Delta)/2$  which is spin split both by the mean-field treatment of $g \sum_{j}  \hat{\vec{S}}_j \cdot \hat{\vec{s}}_j$ interaction term and $\Delta$. The left semi-infinite lead~\cite{Meier2003} for the magnonic subsystem is described by $\hat{H}_\mathrm{m}$ in Eq.~\eqref{eq:hm}, where non-interacting magnons have dispersion \mbox{$\omega_k= 2JS[1-\cos(ka)]+E_ZS$}.

\section{Nonequilibrium diagrammatics for electron-magnon interacting system}\label{sec:diagrammatics}

\subsection{Compact analytical expressions in the Keldysh space}

The NEGF formalism~\cite{Stefanucci2013} operates with two central one-particle quantities---the retarded GF ($\mathbf{G}^r$ for fermions or $\mathbf{B}^r$
for bosons), describing  the density of available quantum states; and the lesser GF ($\mathbf{G}^<$ for fermions or $\mathbf{B}^<$ for bosons),
describing how quasiparticles occupy those states. One can also use two additional GFs, advanced [$\mathbf{G}^a=(\mathbf{G}^r)^\dagger$ for fermions or
$\mathbf{B}^a=(\mathbf{B}^r)^\dagger$ for bosons] and greater ($\mathbf{G}^>$ for fermions or $\mathbf{B}^>$ for bosons), describing the properties of the
corresponding empty states. These four GFs, which generally depend on two time arguments $(t,t^\prime)$, are connected by the fundamental
relation
\begin{equation}\label{eq:fundamental}
\mathbf{A}^>-\mathbf{A}^< = \mathbf{A}^r - \mathbf{A}^a,
\end{equation}
for electronic ($\mathbf{A} \equiv \mathbf{G}$) or bosonic ($\mathbf{A} \equiv \mathbf{B}$) GFs.

Besides having a clear physical meaning, these four GFs make it possible to obtain nonequilibrium expectation values of any one-particle observable, such as charge and spin currents that are the focus of our study. However, these four GFs do not have perturbation expansion akin to zero-temperature GFs on the Feynman contour (the real-time axis from $-\infty$ to $\infty$) or finite-temperature GFs on the Matsubara contour (a segment on the imaginary-time axis from $-i\beta$ to $i\beta$, where $\beta=1/k_BT$). Instead, perturbation expansion is formulated for the contour-ordered~\cite{Stefanucci2013} GF whose two time arguments are located on the Keldysh-Schwinger contour consisting of two counter-propagating copies of the real-time axis---the forward branch extending from $-\infty$ to $\infty$ and the backward branch extending from $\infty$ to $-\infty$. Equivalently, one can introduce $2 \times 2$ matrix GF in the so-called Keldysh space~\cite{Kita2010} which depends on the two time arguments located on the {
\em single} real-time axis extending from $-\infty$ to $\infty$. Such Keldysh-space matrix GF for fermions is defined by
\begin{widetext}
\begin{equation}\label{eq:eGreen}
 \check{\bold{G}}=\left( \begin{array}{cc}
\mathbf{G}^< + \mathbf{G}^r & \mathbf{G}^> \\
\mathbf{G}^< & \mathbf{G}^> - \mathbf{G}^r \end{array} \right)=i\left( \begin{array}{cc}
\theta_{t'-t}\langle \hat{c}^{\dagger}_{1'} \hat{c}_1 \rangle- \theta_{t-t'} \langle \hat{c}_{1} \hat{c}^{\dagger}_{1'} \rangle & -\langle \hat{c}_{1} \hat{c}^{\dagger}_{1'} \rangle \\
\langle \hat{c}^{\dagger}_{1'} \hat{c}_{1} \rangle &  \theta_{t'-t}\langle \hat{c}_{1} \hat{c}^{\dagger}_{1'} \rangle - \theta_{t-t'} \langle \hat{c}^\dagger_{1'} \hat{c}_{1} \rangle\end{array} \right),
\end{equation}
and for bosons it is defined by
\begin{equation}\label{eq:mGreen}
 \check{\bold{B}}=\left( \begin{array}{cc}
\mathbf{B}^< + \mathbf{B}^r & \mathbf{B}^> \\
\mathbf{B}^< & \mathbf{B}^> - \mathbf{B}^r \end{array} \right) = -i \left( \begin{array}{cc}
\theta_{t'-t}\langle \hat{b}^{\dagger}_{1'} \hat{b}_1 \rangle + \theta_{t-t'} \langle \hat{b}_{1} \hat{b}^{\dagger}_{1'} \rangle & \langle \hat{b}_{1} \hat{b}^{\dagger}_{1'} \rangle \\
\langle \hat{b}^{\dagger}_{1'} \hat{b}_{1} \rangle & \theta_{t'-t} \langle \hat{b}_{1} \hat{b}^{\dagger}_{1'} \rangle + \theta_{t-t'} \langle \hat{b}^{\dagger}_{1'} \hat{b}_{1} \rangle \end{array} \right).
\end{equation}
Here $1 \equiv (t,i,\sigma)$ and $1' \equiv (t',j,\sigma')$; $\theta_{x}$ is the Heaviside step function ($\theta_{x}=1$ for $x \ge 0$ and $\theta_{x}=0$ for
$x<0$); and $\langle \cdots \rangle \equiv \mathrm{Tr}[\mathcal{T}_C \cdots \hat{\rho}_0]/\mathrm{Tr}[\hat{\rho}_0]$ is the nonequilibrium expectation value~\cite{Stefanucci2013} with $\mathcal{T}_C$ being the contour ordering operator and the initial density matrix of the system $\hat{\rho}_0$ is usually taken at $-\infty$ for the steady-state formulations within the NEGF. The Keldysh-space matrices (such as $\check{\bold{A}} \equiv \check{\bold{G}}$ or $\check{\bold{A}} \equiv \check{\bold{B}}$) satisfy
\begin{eqnarray}
\check{\bold{A}}^{\dagger} & = & - {\bm \tau}^x \check{\bold{A}} {\bm \tau}^x,
\end{eqnarray}
where we use ${\bm \tau}^{x,y,z}$ to denote the Pauli matrices acting in the Keldysh space.

In stationary problems GFs depend only on the time difference $t-t^\prime$, so that they can be Fourier transformed to energy $E$ or frequency $\omega$.
Using Hamiltonian in Eq.~\eqref{eq:hamiltonian} and performing such Fourier transform leads to the following Keldysh-space Dyson equation for electrons
\begin{equation}\label{eq:eDyson}
\check{\bold{G}}(E)={\bm \tau}^z\frac{1}{(E-\mathbf{H}_\mathrm{e}) {\bm \tau}^z-\check{\bold{\Sigma}}_\mathrm{e-m}(E)-\check{\bold{\Sigma}}_\mathrm{leads}(E)} {\bm \tau}^z,
\end{equation}
or magnons
\begin{equation}\label{eq:mDyson}
\check{\bold{B}}(\omega)={\bm \tau}^z\frac{1}{(\omega-\mathbf{H}_\mathrm{m}) {\bm \tau}^z-\check{\bold{\Omega}}_\mathrm{m-e}(\omega)-\check{\bold{\Omega}}_\mathrm{leads}(\omega)} {\bm \tau}^z.
\end{equation}
\end{widetext}

This approach allows for compact notation by avoiding the widespread route~\cite{Stefanucci2013} where one starts from the Dyson equation for the contour-ordered GF containing convolution integrals on the two-branch Keldysh-Schwinger contour, and then applies the so-called Langreth rules~\cite{Stefanucci2013} to find the lengthy expressions involving the lesser and retarded GFs with two time arguments located on the single real-time axis (or their Fourier transforms). The Dyson equation for the Keldysh-space matrix GFs in energy, like Eqs.~\eqref{eq:eDyson} and ~\eqref{eq:mDyson}, is rarely used in the literature due to redundancy expressed by Eq.~\eqref{eq:fundamental}. For example, such equation can be found in the NEGF-based calculations of the full counting statistics~\cite{Gogolin2006,Urban2010,Novotny2011} where the presence of the counting field results in the nonunitary evolution on the Keldysh-Schwinger contour, thereby requiring to work with all four submatrices in Eqs.~\eqref{eq:eGreen} or ~\eqref{eq:mGreen} because the relation like Eq.~\eqref{eq:fundamental} is not valid anymore. Nevertheless, even when Eq.~\eqref{eq:fundamental} holds, it can be advantageous to work directly with the matrices in the Keldysh space---this greatly simplifies writing the analytical expressions that the Feynman diagrams of nonequilibrium MBPT represent and, moreover, it makes possible to derive expressions for the perturbation expansion of more complicated quantities like the current noise obtained from the two-particle nonequilibrium correlation function.~\cite{Mahfouzi2013b}

\begin{figure*}
\includegraphics[scale=0.6,angle=0]{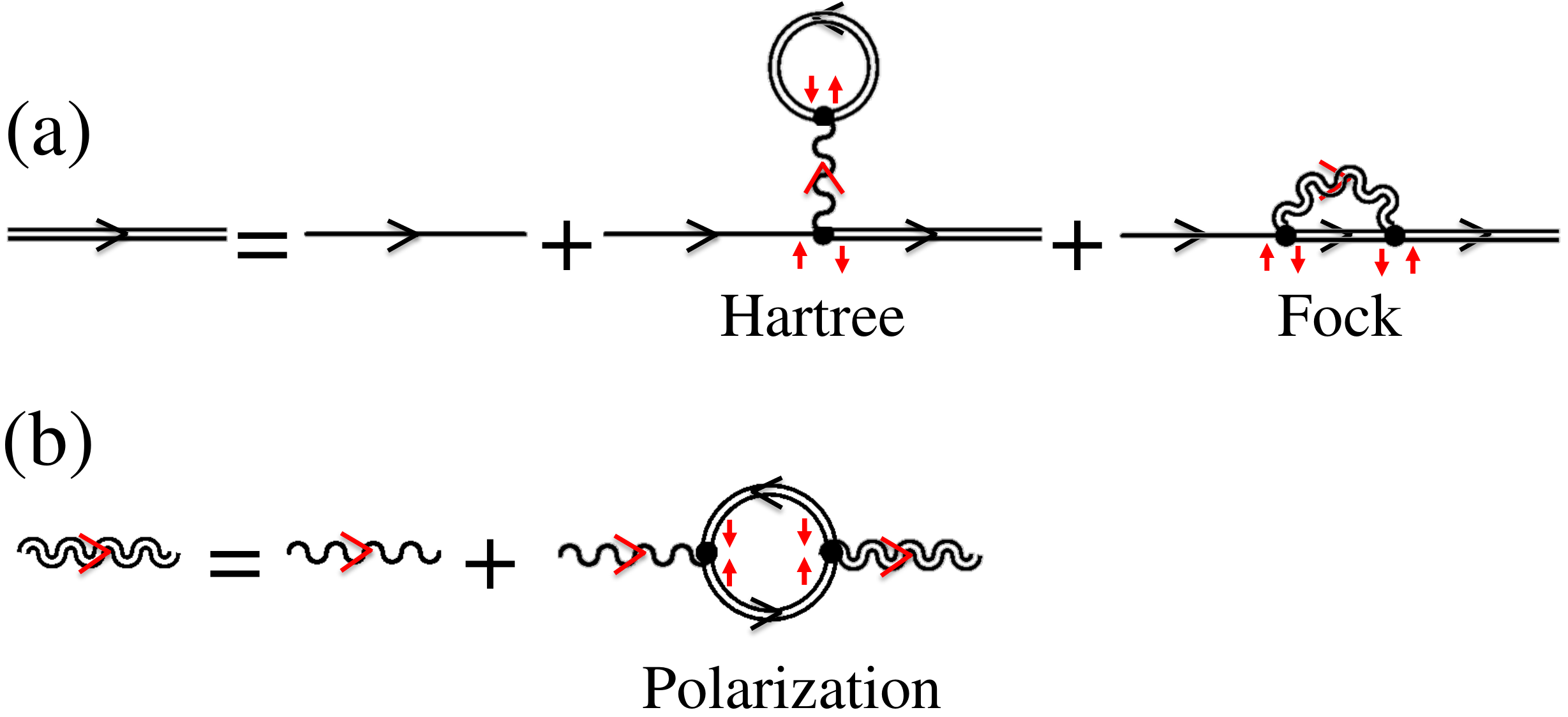}
\caption{(Color online) Diagrammatic representation of the Dyson equation in the Keldysh space: (a) the electron case, $\check{\bold{G}} = \check{\bold{G}}_0 + \check{\bold{G}}_0 \check{\bm \Sigma}_\mathrm{e-m} \check{\bold{G}}$, in Eq.~\eqref{eq:eDyson}; and (b) the magnon case, $\check{\bold{B}} = \check{\bold{B}}_0 + \check{\bold{B}}_0 \check{\bm \Omega}_\mathrm{m-e}\check{\bold{B}}$, in Eq.~\eqref{eq:mDyson}. The perturbation expansion for the electronic self-energy $\check{\bm \Sigma}_\mathrm{e-m}$ in (a) retains Hartree and Fock diagrams, while the expansion in (b) for the magnonic self-energy
$\check{\bm \Omega}_\mathrm{m-e}$ retains e-h polarization bubble diagram. The single straight line denotes the non-interacting electronic GF, $\check{\bold{G}}_0$ [which includes the self-energies due to the leads in Eq.~\eqref{eq:eKSelf}]; the single wavy line denotes the non-interacting magnonic GF, $\check{\bold{B}}_0$ [which includes the self-energy due to the left lead in Eq.~\eqref{eq:mKSelf}];  double straight line denotes the interacting electronic GF, $\check{\bold{G}}$; and double wavy line denotes the interacting magnonic GF, $\check{\bold{B}}$. The solid circles denote vertices that are integrated out. The electron spin is flipped at each vertex, which is illustrated by spin-$\uparrow$ (before the vertex) being flipped into spin-$\downarrow$ (after the vertex), while a magnon is being created. The same process applies to flipping of spin-$\downarrow$ to spin-$\uparrow$, where the direction of magnon propagation (indicated by arrow on the wavy lines) is reversed. Note that the Hartree diagram in panel (a) contains a single (rather than double) wavy line in order to avoid double counting.}
\label{fig:fig2}
\end{figure*}

The self-energies due to many-body interaction, $\check{\bm \Sigma}_\mathrm{e-m}(E)$ for electron and $\check{\bm \Omega}_\mathrm{m-e}(\omega)$ for magnon, can be simply added to the self-energies introduced by the attached semi-infinite leads, $\check{\bm \Sigma}_\mathrm{leads}$ for electron and $\check{\bm \Omega}_\mathrm{leads}(\omega)$ for magnons, respectively. This is due to the fact that \mbox{e-m} interaction is assumed to be localized within the active region in Fig.~\ref{fig:fig1}, so that the leads do not involve many-body interactions. Thus, the self-energies of the leads for the junction in Fig.~\ref{fig:fig1}
\begin{widetext}
\begin{align}
&\check{\bold{\Sigma}}_\mathrm{leads}=\sum_{\alpha=L,R}\left( \begin{array}{cc}
(1-f_{\alpha}){\bm \Sigma}^r_{\alpha}+f_{\alpha}{\bm \Sigma}^a_{\alpha} & (1-f_{\alpha})({\bm \Sigma}^r_{\alpha}-{\bm \Sigma}^a_{\alpha}) \\
-f_{\alpha}({\bm \Sigma}^r_{\alpha}-{\bm \Sigma}^a_{\alpha}) & -f_{\alpha}{\bm \Sigma}^r_{\alpha}-(1-f_{\alpha}){\bm \Sigma}^a_{\alpha} \end{array} \right)\label{eq:eKSelf}\\
&\check{\bold{\Omega}}_\mathrm{leads}=\sum_{\alpha=L}\left( \begin{array}{cc}
(1+n_{\alpha}){\bm \Omega}^r_{\alpha}-n_{\alpha}{\bm \Omega}^a_{\alpha} & (1+n_{\alpha})({\bm \Omega}^r_{\alpha} - {\bm \Omega}^a_{\alpha}) \\
n_{\alpha}({\bm \Omega}^r_{\alpha} - {\bm \Omega}^a_{\alpha}) & n_{\alpha} {\bm \Omega}^r_{\alpha}-(1+n_{\alpha}) {\bm \Omega}^a_{\alpha} \end{array} \right)\label{eq:mKSelf}
\end{align}
\end{widetext}
are single-particle quantities which can always be computed in an exact fashion, either analytically~\cite{Datta1995} for simple models like Eqs.~\eqref{eq:he} and ~\eqref{eq:hm} or numerically for more complicated lead Hamiltonians.~\cite{Rungger2008} The effect of the bias voltage is introduced by a rigid shift in energy, ${\bm \Sigma}^r_{L,R}(E,V_b)= {\bm \Sigma}^r_{L,R}(E \mp eV_b/2,0)$. The Fermi function of the macroscopic reservoir to which lead $\alpha$ is attached is denoted by $f_\alpha(E)=f(E - eV_\alpha)$. The Bose-Einstein distribution function of the macroscopic reservoir of magnons to which the left F layer is attached is denoted by $n_\alpha(\omega)$.

The one-particle self-energies due to \mbox{e-m} interaction, $\check{\bm \Sigma}_\mathrm{e-m}$ and magnon $\check{\bm \Omega}_\mathrm{m-e}$, are formally obtained by summing all irreducible diagrams, i.e., those diagrams that cannot be taken apart by cutting a single line. The self-energies are actually {\em functionals} of the respective electronic or magnonic GF, so that they have to be approximated in practical calculations. Diagrammatic techniques provide a natural scheme for generating approximate self-energies, as well as for systematically improving these approximations. While there are no general prescriptions on how to select the relevant diagrams, this process can be guided by physical intuition. In addition, unlike equilibrium~\cite{Richmond1970} MBPT, the diagrams selected in nonequilibrium~\cite{Stefanucci2013} MBPT must generate GFs that yield expectation value for  charge current which is conserved. For example, the final current in the left and right leads of the device in Fig.~\ref{fig:fig1} must be the same in any chosen approximation for the self-energies.

Here we select one of the conserving approximations,~\cite{Stefanucci2013} where the Feynman diagrams retained for the electron or magnon self-energies are displayed in Figs.~\ref{fig:fig2}(a) and ~\ref{fig:fig2}(b), respectively. The diagrams in Figs.~\ref{fig:fig2}(a) are equivalent to the so-called self-consistent Born approximation (SCBA) for the electronic self-energy considered in problems like \mbox{e-ph} interacting systems.~\cite{Mitra2004,Viljas2005,Koch2006,Frederiksen2007,Galperin2007,Lu2007,Asai2008,Lee2009a,Dash2011} However, in the case of \mbox{e-m} interaction, one has to introduce additional bookkeeping in such diagrams to account for flipping of electron spin together with magnon emission or absorption, as illustrated in Fig.~\ref{fig:fig2}. The Keldysh-space expression for the electronic self-energy read from the Fock diagram in Fig.~\ref{fig:fig2}(a) is given by
\begin{subequations}\label{eq:fock}
\begin{align}
&\check{\bold{\Sigma}}_{mn,\uparrow\uparrow}^{F}(E)=\frac{i}{2} g^2 S \int d\omega \, \check{\bold{B}}_{mn}(\omega)\check{\bold{G}}_{mn,\downarrow\downarrow}(E-\omega), \\
&\check{\bold{\Sigma}}_{mn,\downarrow\downarrow}^{F}(E)=\frac{i}{2} g^2 S \int d\omega\, [\check{\bold{B}}^T]_{mn}(\omega)  \check{\bold{G}}_{mn,\uparrow\uparrow}(E+\omega).
\end{align}
\end{subequations}
Here $m$ and $n$ are matrix indices which include the Keldysh space and real (i.e., orbital) space, so that $\check{\bold{\Sigma}}_{mn,\sigma\sigma}^{F}(E)$ selects a submatrix of $\check{\bold{\Sigma}}^F$.

Finding the proper expression for the Hartree diagram in Fig.~\ref{fig:fig2}(a) in the Keldysh space requires extra care~\cite{Oguri2013} because $t$ and $t^\prime$ for the inner electronic GF along the loop are equal, so that $\theta_{t-t'}$ in Eq.~\eqref{eq:eGreen} gives 1 or 0 depending on whether
$t-t' \rightarrow 0^+$ or $t-t' \rightarrow 0^-$, respectively. Therefore, $\check{\bf G}(t,t' \rightarrow t+0^+) = \left( \begin{array}{cc} \mathbf{G}^<(t,t) &  0 \\ 0 & \mathbf{G}^<(t,t) \end{array}\right)$ in terms of which we obtain the following expressions
\begin{subequations}\label{eq:hartree}
\begin{align}
&\check{\bm \Sigma}_{mn,\uparrow\downarrow}^{H} = \frac{-i\delta_{mn}}{2} g^2S  \sum_p \int dE \,  [\check{\bf B}_0]_{mp}(\omega=0) {\bm \tau}^z_{pp} \check{\bf G}_{pp,\downarrow \uparrow }(E), \\
&\check{\bm \Sigma}_{mn,\downarrow\uparrow}^{H} = \frac{-i\delta_{mn}}{2} g^2S  \sum_p \int dE  \,  [\check{\bf B}^T_0]_{mp}(\omega=0) {\bm \tau}^z_{pp} \check{\bf G}_{pp, \uparrow\downarrow}(E).
\end{align}
\end{subequations}
Note that the off-diagonal (i.e., lesser and greater) components of $\check{\bold{\Sigma}}_\mathrm{m-e}^H$ vanish, and the remaining retarded component on the diagonal is energy independent.

Although  \mbox{$\check{\bold{\Sigma}}_\mathrm{e-m}(E) = \check{\bold{\Sigma}}^H + \check{\bold{\Sigma}}^F(E)$} in SCBA, we retain only the Fock term in the actual calculations below. We note that in SCBA for \mbox{e-ph} interacting systems, $\check{\bold{\Sigma}}^H$ is often neglected~\cite{Frederiksen2007,Lee2009a} due to being small and, therefore, having little effect on the final current (this becomes unwarranted for larger \mbox{e-ph} interaction strengths where SCBA breaks down~\cite{Lee2009a}). For the \mbox{e-m} interacting systems, the situation is much more complex because direct evaluation of Eq.~\eqref{eq:hartree} leads to numerical instabilities. This stems from the fact that our MTJ is invariant with respect to the rotation around the $z$-axis (see Fig.~\ref{fig:fig1}), so that spin-flip rate which appears in Eq.~\eqref{eq:hartree} can acquire arbitrary phase thereby requiring to
consider full double time dependence of $\check{\bold{\Sigma}}^H$. We relegate this to future studies, while here we retain $\check{\bold{\Sigma}}_\mathrm{e-m}(E) = \check{\bold{\Sigma}}^F(E)$ which is termed~\cite{Lee2009a} Fock-only SCBA (F-SCBA).

In the case of \mbox{e-ph} many-body systems driven far from equilibrium, phonon heating due to propagating electrons has been
considered either phenomenologically using a rate equation for the phonon occupation,~\cite{Koch2006,Frederiksen2007,Siddiqui2007} or
microscopically by using phonon GF with interacting self-energy truncated to the e-h
polarization bubble diagram.~\cite{Mitra2004,Viljas2005,Galperin2007,Lu2007,Asai2008,Urban2010,Novotny2011} It is worth mentioning that
the two approaches yield virtually identical results for time-averaged current in the limit of weak \mbox{e-ph} coupling, but they start
differing significantly in the case of the current noise due to the feedback of the phonon dynamics on the statistics of the transmitted electrons which cannot be captured by the phenomenological rate equation approach.~\cite{Novotny2011} Since magnon bandwidth ($\simeq 100$ meV) is relatively small,~\cite{Balashov2008} they can be easily driven into far from equilibrium state by charge current at finite bias voltage. For the purpose of
describing such state, we retain in Fig.~\ref{fig:fig2}(b) the e-h polarization bubble diagram for the magnonic self-energy whose analytical expression
is given by
\begin{equation}\label{eq:bubble}
\check{\bold{\Omega}}^\mathrm{pol}_{mn}(\omega)=-\frac{i}{2} g^2S \int dE\, \check{\bold{G}}_{\downarrow\downarrow,nm}(E) \check{\bold{G}}_{\uparrow\uparrow,mn}(E+\omega).
\end{equation}
Thus, the dressed magnonic GF which includes this self-energy through Eq.~\eqref{eq:mDyson} will be inserted into the electronic self-energy diagrams in Fig.~\ref{fig:fig2}(a), thereby generating an infinite resummation of diagrams until the mutual self-consistency is achieved. Note that this is analogous to
the self-consistent GW treatment of the one-particle electronic self-energy due to electron-electron interaction out of equilibrium.~\cite{Thygesen2008}

\subsection{Numerical implementation in Keldysh space}\label{sec:numerics}

Equations~\eqref{eq:eDyson}, ~\eqref{eq:mDyson}, ~\eqref{eq:fock} and ~\eqref{eq:bubble} form a system of coupled nonlinear integral equations
that has to be solved iteratively until the convergence criterion is met. We use expectation value of charge current (see Sec.~\ref{sec:zba}) to
define one such criterion, $\sum_{\alpha=L,R} |I_\alpha^\mathrm{new} - I_\alpha^\mathrm{old}| < \delta$. Here $I_\alpha^\mathrm{old}$ is charge current
in lead $\alpha$ at the beginning of an iteration, $I_\alpha^\mathrm{new}$ denotes charge current at the end of the same iteration and we select $\delta=10^{-6}$.

\begin{figure}
\includegraphics[scale=0.25,angle=0]{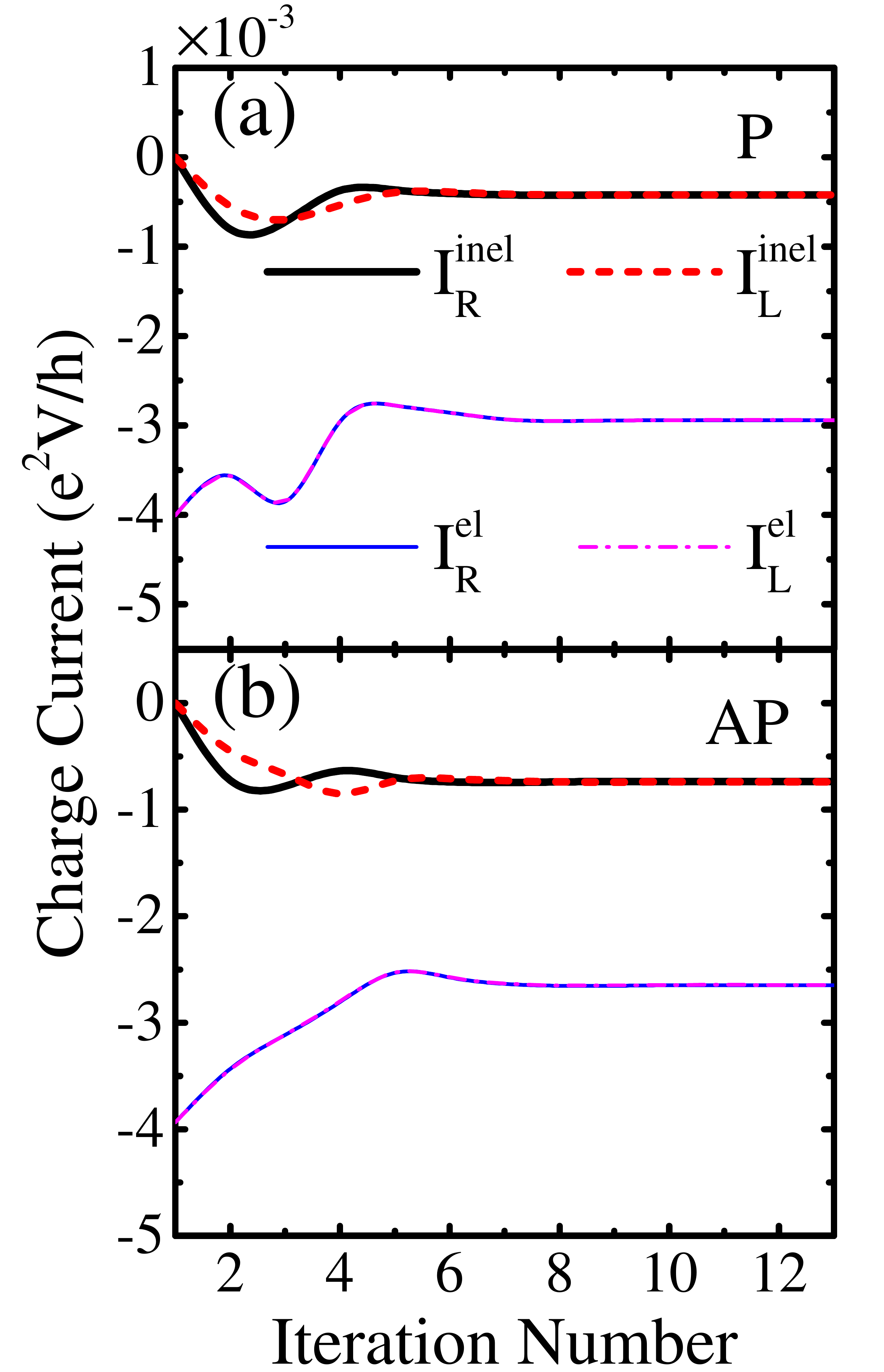}
\caption{(Color online) Elastic and inelastic contributions to charge current in the left and right lead of MTJ in Fig.~\ref{fig:fig1} with $N_x \times N_y \equiv 3 \times 1$ active region at different iteration number within the self-consistent loop for solving coupled Eqs.~\eqref{eq:eDyson}, ~\eqref{eq:mDyson}, ~\eqref{eq:fock} and ~\eqref{eq:bubble}. The orientation of the magnetizations of two F layers in Fig.~\ref{fig:fig1} is parallel in (a) and antiparallel in (b). The bias voltage is set as \mbox{$V_b = -60$ mV} and temperature is $T=12$ K.}
\label{fig:fig3}
\end{figure}

The Keldysh-space electronic GF and self-energies in these coupled equations are matrices of the size
\mbox{$N_\mathrm{sites} \times 2_\mathrm{spin} \times 2_\mathrm{Keldysh}$} (if $n_\mathrm{orb}>1$ orbitals are used per site then
\mbox{$N_\mathrm{sites} \mapsto N_\mathrm{sites} \times n_\mathrm{orb}$}), while the magnonic GF and self-energies are matrices of the size
\mbox{$N_\mathrm{sites} \times 2_\mathrm{Keldysh}$}. The most time-consuming part of solving the coupled equations is the integration in Eq.~\eqref{eq:fock} for $\check{\bm \Sigma}^F(E)$, which can be viewed as the convolution
\begin{equation}\label{eq:convolution}
\check{\bold C}(x) = \check{\bold A} (x) \ast \check{\bold B}(x)= \int\limits_{-\infty}^{\infty}\!\! dy\, \check{\bold A}(x-y) \circ \check{\bold B}(y),
\end{equation}
where $[\check{\bold A} \circ  \check{\bold B}]_{mn} =[\check{\bold A}]_{mn}  [\check{\bold B}]_{mn}$ is the elementwise product of matrices.
The fact that matrix elements of the retarded and advanced components of GFs in Eqs.~\eqref{eq:eDyson} and ~\eqref{eq:mDyson} are nonzero in the whole
range of integration would make the numerical computation of this convolution prohibitively expensive. However, this obstacle can be removed by using the fact that retarded and advanced GFs are analytic functions in the upper and lower half of the complex plane, respectively. Thus, the real ($\Re$) and imaginary ($\Im$) parts of their matrix elements  must obey the following relation
\begin{eqnarray}\label{eq:ht}
\Re\{\mathbf{A}^r(x) \} = \mathcal{H}[\Im\{\mathbf{A}^r(x)\}] = \frac{1}{\pi} \, \mathcal{P} \int dy\, \frac{\Im \{\mathbf{A}^r(y)\} }{x-y}.
\end{eqnarray}
Here $\mathcal{H}$ is the Hilbert transform, whose implementation in our scheme is discussed in more details in Appendix~\ref{sec:ap_hilbert}, and $\mathcal{P}$ stands for the Cauchy principal value. This makes it possible to decompose Keldysh-space matrices as follows
\begin{align}
 \check{\bold{A}}&=\left( \begin{array}{cc}
\mathbf{A}_{11} & \mathbf{A}_{12} \\
\mathbf{A}_{21} & \mathbf{A}_{22} \end{array} \right)=\left( \begin{array}{cc}
\mathbf{A}^<+\mathbf{A}^r & \mathbf{A}^> \\
\mathbf{A}^< & \mathbf{A}^>-\mathbf{A}^r \end{array} \right)\nonumber\\
&=\check{\bold{A}}^\mathrm{sym}+\mathcal{H}[\check{\mathcal{\bold{A}}}^\mathrm{asym}],
\end{align}
where
\begin{subequations}
\begin{align}
&\check{\mathcal{\bold{A}}}^\mathrm{sym}=\left( \begin{array}{cc}
\Re(\mathbf{A}_{21})+i\Im(\mathbf{A}_{11}) & \mathbf{A}_{12} \\
\mathbf{A}_{21} & \Re(\mathbf{A}_{12})+i\Im(\mathbf{A}_{22}) \end{array} \right), \\
&\check{\bold{A}}^\mathrm{asym}=\left( \begin{array}{cc}
1 & 0 \\
0 & -1 \end{array} \right)\Im(\mathbf{A}_{11} - \mathbf{A}_{21}),
\end{align}
\end{subequations}
are labeled as ``symmetric'' and ``asymmetric'' part. We note that all the relevant information is already
contained in $\check{\bold{A}}^\mathrm{sym}$, so that one can find $\check{\bold{A}}^\mathrm{asym}$ from it by using
\begin{align}\label{eq:asymtosym}
\check{\bold{A}}^\mathrm{asym}=\frac{1}{i} \left( \begin{array}{cc}
1 & 0 \\
0 & -1 \end{array} \right)(\mathbf{A}^\mathrm{sym}_{11}-\mathbf{A}^\mathrm{sym}_{21}).
\end{align}
This idea allows us to restrict the range of integration in the convolution in Eq.~\eqref{eq:convolution} to the energy bandwidth of electrons and magnons.
Once the decomposition is done for both matrices $\check{\bold A}$ and $\check{\bold B}$, the convolution in Eq.~\eqref{eq:convolution} can be evaluated using
\begin{align}
\check{\mathcal{\bold{C}}}= \check{\bold{A}} \ast \check{\bold{B}} = \check{\bold{C}}^\mathrm{sym} + \mathcal{H}[\check{\bold{C}}^\mathrm{asym}],
\end{align}
with
\begin{subequations}
\begin{align}
& \check{\bold{C}}^\mathrm{sym} = \check{\bold{A}}^\mathrm{sym} \ast \check{\bold{B}}^\mathrm{sym} - \check{\bold{A}}^\mathrm{asym} \ast \check{\bold{B}}^\mathrm{asym}, \\
& \check{\bold{C}}^\mathrm{asym}= \check{\bold{A}}^\mathrm{sym} \ast \check{\bold{B}}^\mathrm{asym} + \check{\bold{A}}^\mathrm{asym} \ast \check{\bold{B}}^\mathrm{sym}.
\end{align}
\end{subequations}
Here we used the following properties of the Hilbert transform and the convolution operator
\begin{align}
&\mathcal{H}[\check{\bold{A}}] \ast \check{\bold{B}}=\check{\bold{A}} \ast \mathcal{H}[\check{\bold{B}}] = \mathcal{H}[\check{\bold{A}} \ast \check{\mathcal{\bold{B}}}], \\
&\mathcal{H}[\mathcal{H}[\check{\bold{A}}]]=-\check{\bold{A}}.
\end{align}
Note that one has to actually calculate only $\check{\bold{C}}^\mathrm{sym}$, after which the asymmetric part is obtained from
Eq.~\eqref{eq:asymtosym}.

\begin{figure}
\includegraphics[scale=0.36,angle=0]{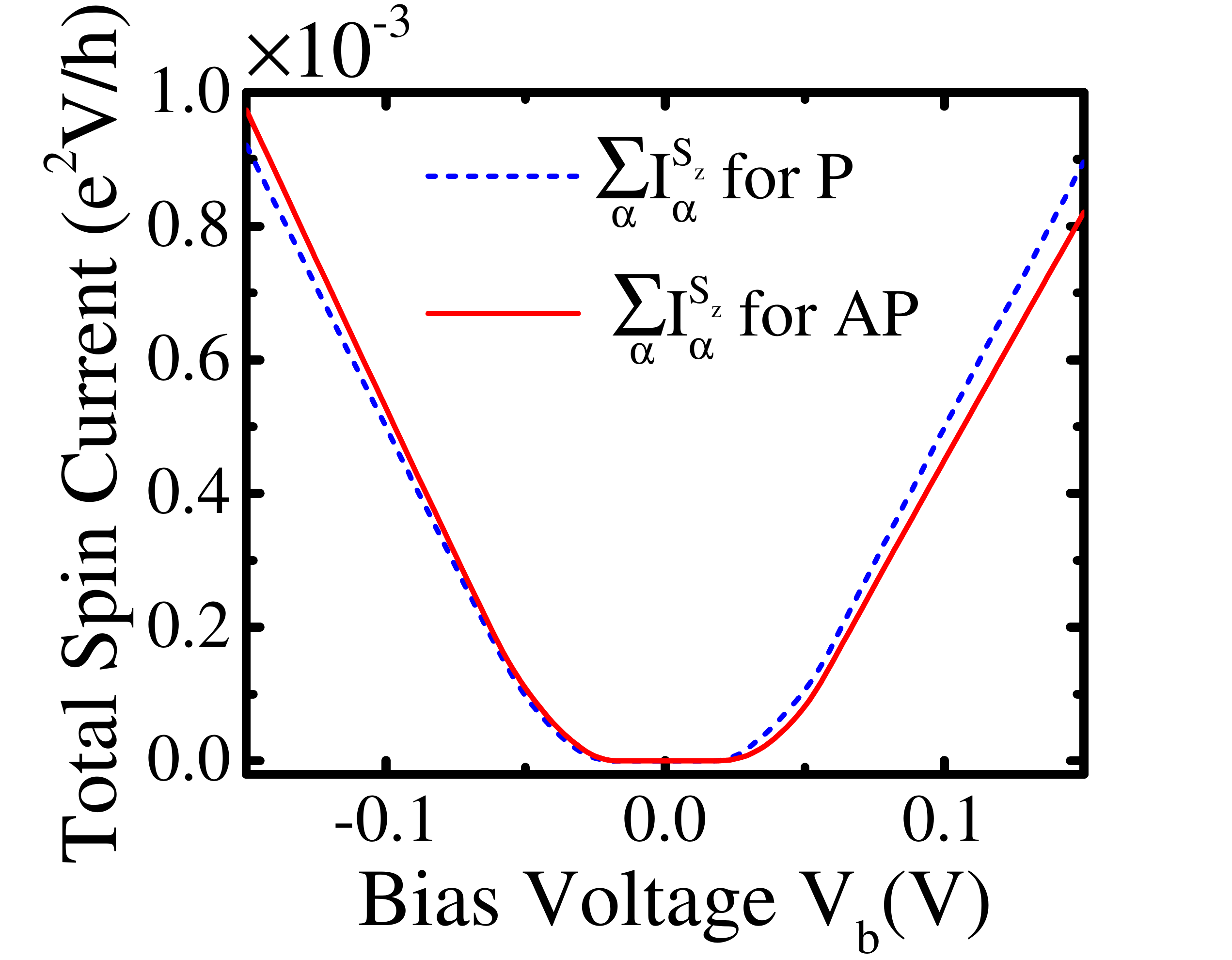}
\caption{(Color online) Total spin current dissipated (at \mbox{$T=12$ K}) inside $N_x \times N_y \equiv 3 \times 1$ active region of MTJ in Fig.~\ref{fig:fig1} as a function of the bias voltage  for parallel and antiparallel orientation of the magnetizations of two F layers. The spin current is obtained from Eq.~\eqref{eq:landauerspin} using electronic GF computed by solving coupled Eqs.~\eqref{eq:eDyson}, ~\eqref{eq:mDyson} and ~\eqref{eq:fock}.}
\label{fig:ISV}
\end{figure}

\section{Application to charge and spin currents in MTJ driven by finite bias voltage}\label{sec:zba}

The charge current in lead $\alpha$ can be viewed as the sum of two spin-resolved charge currents, $I_\alpha = I_\alpha^\uparrow + I_\alpha^\downarrow$. For an interacting active region attached to two non-interacting semi-infinite leads, lead currents can be obtained directly from $\check{\bold G}(E)$ and $\check{\bold{\Sigma}}_{\alpha}$
\begin{align}\label{eq:mw}
I_{\alpha}&=\frac{e}{h} \int dE \, \mathrm{Tr} \left[{\bm \tau}_{11} (\check{\bold{G}} {\bm \tau}^z \check{\bold{\Sigma}}_{\alpha}-\check{\bold{\Sigma}}_{\alpha} {\bm \tau}^z \check{\bold{G}})\right]  \nonumber \\
& = \frac{e}{h} \int dE\, \mathrm{Tr} \left[ \mathbf{G}^>(E) {\bm \Sigma}^<_{\alpha}(E) - \mathbf{G}^<(E) {\bm \Sigma}^>_{\alpha}(E) \right],
\end{align}
where we employ the following notation
\begin{equation}
 {\bm \tau}_{11}=\left( \begin{array}{cc}
 1 & 0 \\
0 & 0
\end{array} \right).
\end{equation}
The second line in Eq.~\eqref{eq:mw} is the well-known Meir-Wingreen formula.~\cite{Meir1992} Similarly, the spin current  \mbox{$I_\alpha^S = I_\alpha^\uparrow - I_\alpha^\downarrow$} (in the same units as for the charge current) in lead $\alpha$ is obtained from
\begin{align}\label{eq:mws}
I_{\alpha}^{S_{x,y,z}} & = \frac{e}{h} \int dE \,\mathrm{Tr} \left[ {\bm \tau}_{11} {\bm \sigma}^{x,y,z}(\check{\bold{G}}{\bm \tau}^z \check{\bold{\Sigma}}_{\alpha} - \check{\bold{\Sigma}}_{\alpha}{\bm \tau}^z \check{\bold{G}})\right] \nonumber \\
& = \frac{e}{h} \int dE\, \mathrm{Tr} \left[ {\bm \sigma}^{x,y,z} \left( \mathbf{G}^>(E) {\bm \Sigma}^<_{\alpha}(E) - \mathbf{G}^<(E){\bm \Sigma}^>_{\alpha}(E) \right) \right].
\end{align}

\begin{figure}
\includegraphics[scale=0.38,angle=0]{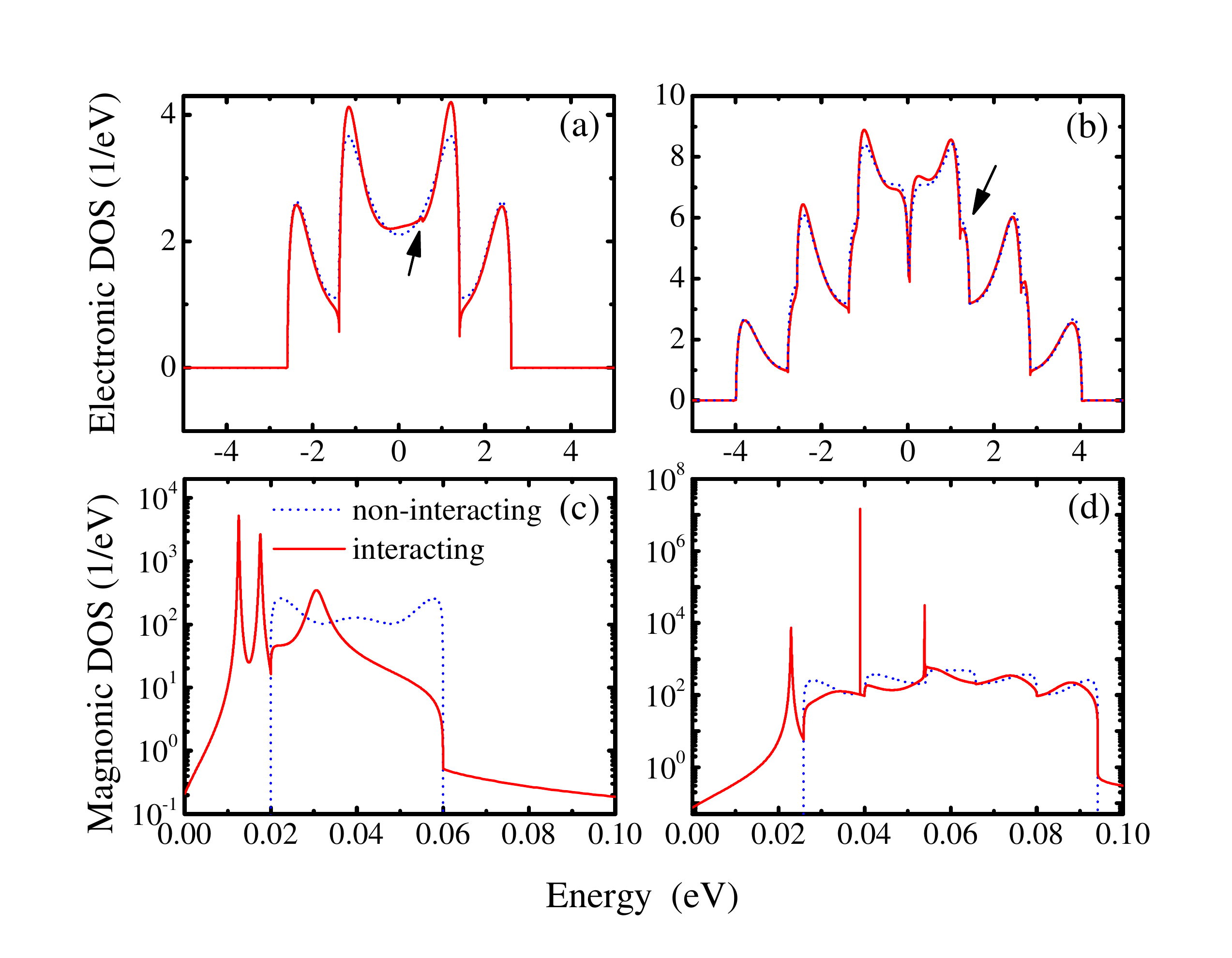}
\caption{(Color online) The electronic density of states within the active region of MTJ model in Fig.~\ref{fig:fig1} of size: (a) $N_x \times N_y \equiv 3 \times 1$; and (b) $N_x \times N_y \equiv 3 \times 3$. The magnonic density of states within the same active regions is shown in panels (c) and (d), respectively. These quantities are computed at finite bias voltage \mbox{$V_b = -60$ mV} and at temperature \mbox{$T=12$ K}, in the absence ($g=0$ for dashed line) or the presence ($g \neq 0$ for solid line) of \mbox{e-m} interaction in the Hamiltonian in Eq.~\eqref{eq:hem}. The respective DOS is obtained from the retarded electronic GF, using $-\mathrm{Tr}[\Im(\mathbf{G}^r)]/\pi$, or the retarded magnonic GF, using $-\mathrm{Tr}[\Im(\mathbf{B}^r)]/\pi$, after solving coupled Eqs.~\eqref{eq:eDyson}, ~\eqref{eq:mDyson}, ~\eqref{eq:fock} and ~\eqref{eq:bubble} which take into account influence of electrons on magnons. The arrows in panels (a) and (b) point at the kinks (located at the Fermi energies of MTJ model with two different sizes of the active region, respectively) in the interacting electronic DOS (solid line) due to e-m coupling.}
\label{fig:fig5}
\end{figure}
\begin{figure*}
\includegraphics[scale=0.6,angle=0]{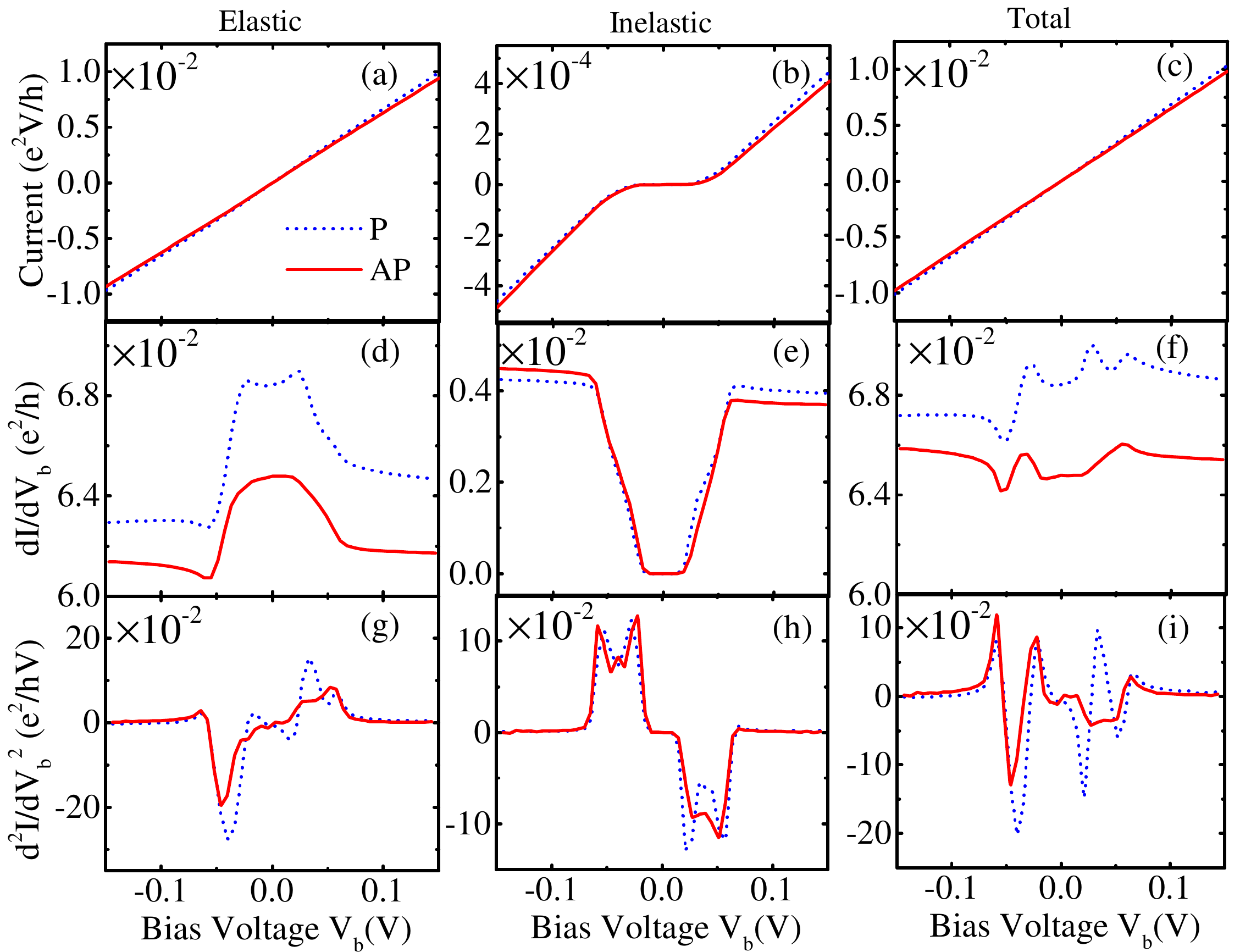}
\caption{(Color online) Elastic and inelastic charge currents in Eq.~\eqref{eq:landauer} and their sum, as well as the corresponding first and second derivatives, versus the bias voltage in the model of MTJ in Fig.~\ref{fig:fig1} with active region $N_x \times N_y \equiv 3 \times 1$ for parallel and antiparallel orientation of its magnetizations. These charge currents are obtained from the electronic GF computed in F-SCBA which includes (in self-consistent fashion) the non-interacting magnonic GF. The temperature is set as $T=12$ K.}
\label{fig:fig6}
\end{figure*}

The charge current in Eq.~\eqref{eq:mw} can be conveniently separated~\cite{Asai2008,Ness2010} into two terms, \mbox{$I_{\alpha}=I_{\alpha}^\mathrm{el}+I_{\alpha}^\mathrm{inel}$}
\begin{subequations}\label{eq:landauer}
\begin{align}
&I_{\alpha}^\mathrm{el}=\frac{e}{h}\sum_{\beta}{\int dE\, \mathrm{Tr} \left[ {\bm \Gamma}_{\alpha} \mathbf{G}^r {\bm \Gamma}_{\beta} \mathbf{G}^a\right] \left(f_{\beta}-f_{\alpha}\right)}, \label{eq:el}\\
&I_{\alpha}^\mathrm{inel}=\frac{e}{h} \int\!\! dE\, \mathrm{Tr} \left[ \mathbf{G}^r {\bm \Sigma}^{>,F} \mathbf{G}^a {\bm \Sigma}^<_{\alpha} - \mathbf{G}^r {\bm \Sigma}^{<,F} \mathbf{G}^a {\bm \Sigma}^>_{\alpha} \right]. \label{eq:inel}
\end{align}
\end{subequations}
We label the first term as ``elastic'' current $I_{\alpha}^\mathrm{el}$ since it has the form of the Landauer-like formula for elastic transport of non-interacting quasiparticles whose effective transmission function is expressed~\cite{Caroli1971} in terms of NEGF quantities, \mbox{$T(E) =\mathrm{Tr}[{\bm \Gamma}_{\alpha} \mathbf{G}^r {\bm \Gamma}_{\beta} \mathbf{G}^a]$}. The second term appears as the nonequilibrium corrections due to many-body interactions, which we label as ``inelastic'' current. Plotting separately elastic and inelastic current components makes it possible to provide additional
insights when interpreting our results in Sec.~\ref{sec:zba}.

\begin{figure*}
\includegraphics[scale=0.6,angle=0]{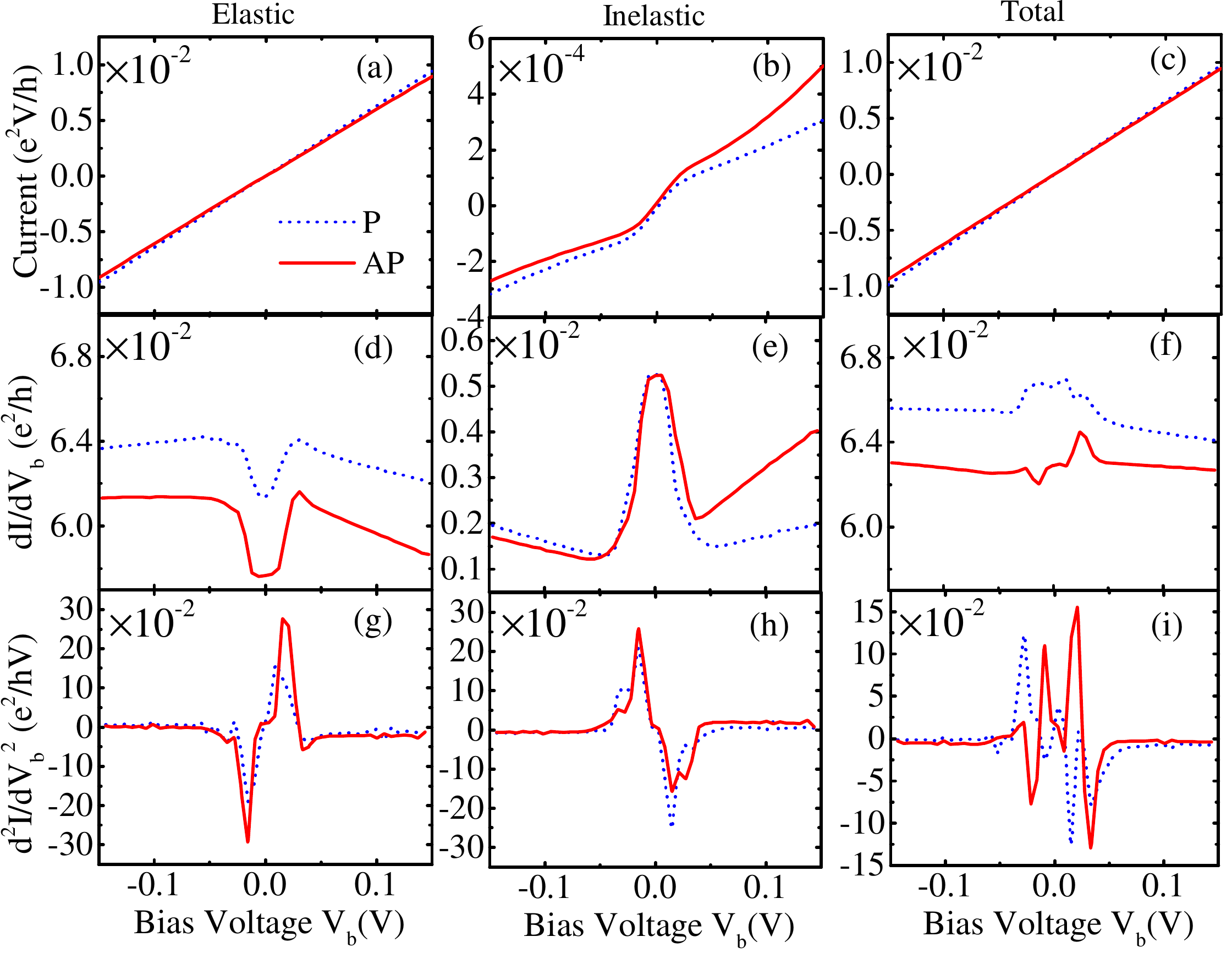}
\caption{(Color online) Elastic and inelastic charge currents in Eq.~\eqref{eq:landauer} and their sum, as well as the corresponding first and second derivatives, versus the bias voltage in the model of MTJ in Fig.~\ref{fig:fig1} with active region $N_x \times N_y \equiv 3 \times 1$ for parallel and antiparallel orientation of its magnetizations. These charge currents are obtained from the electronic GF computed in F-SCBA, which includes (in self-consistent fashion) the interacting magnonic GF with the e-h polarization bubble diagram in Fig.~\ref{fig:fig2}(b). The temperature is set as $T=12$ K.}
\label{fig:fig7}
\end{figure*}

Note that apparent connection of Eq.~\eqref{eq:el} to the Landauer formula should  not be pushed too far since the effective transmission $T(E)$ in $I_{\alpha}^\mathrm{el}$ already contains part of \mbox{e-m} interaction. That is, the standard Landauer formula~\cite{Caroli1971} for single-particle elastic scattering uses the retarded and advanced GFs which include the self-energies due to the semi-infinite leads only. On the other hand, $\mathbf{G}^r$ and $\mathbf{G}^a$ in $I_{\alpha}^\mathrm{el}$ include additional self-energy due to \mbox{e-m} interaction which renormalizes the non-interacting reference
system, and for strong enough interaction can go even beyond the quasiparticle description of the many-body interacting quantum system. Even when 
quasiparticles are well-defined, the presence of self-energy that is functional of the retarded GF itself means that $I_{\alpha}^\mathrm{el}$ includes
dephasing effects due to many-body interaction~\cite{Chen2012} and is, therefore, different from phase-coherent tunneling current that would be obtained  
from the the standard Landauer formula.~\cite{Caroli1971}

Here we illustrate in Fig.~\ref{fig:fig3} that $I_{\alpha}^\mathrm{el}$ is conserved at each iteration, while the conservation of $I_{\alpha}^\mathrm{inel}$ component requires to reach the self-consistency in the computation of electronic GF and self-energy in F-SCBA, as discussed in Eq.~\eqref{eq:proof}. Note that the magnonic GF and self-energy used to obtain Fig.~\ref{fig:fig3} also include e-h polarization bubble diagram from Fig.~\ref{fig:fig2}(b).

The spin current can analogously be separated into the elastic and inelastic contributions, \mbox{$I_\alpha^{S_z} = I_\alpha^{S_{z},\mathrm{el}} + I_\alpha^{S_{z},\mathrm{inel}}$}
\begin{subequations}\label{eq:landauerspin}
\begin{align}
&I_{\alpha}^{S_z,\mathrm{el}}=\frac{e}{h}\sum_{\beta}{\int dE\, \mathrm{Tr} \left[{\bm \sigma}^{z} {\bm \Gamma}_{\alpha} \mathbf{G}^r {\bm \Gamma}_{\beta} \mathbf{G}^a\right] \left(f_{\beta}-f_{\alpha}\right)}, \label{eq:elspin}\\
&I_{\alpha}^{S_z,\mathrm{inel}}=\frac{e}{h} \int\!\! dE\, \mathrm{Tr} \left[ {\bm \sigma}^{z} \left( \mathbf{G}^r {\bm \Sigma}^{>,F} \mathbf{G}^a {\bm \Sigma}^<_{\alpha} - \mathbf{G}^r {\bm \Sigma}^{<,F} \mathbf{G}^a {\bm \Sigma}^>_{\alpha} \right) \right]. \label{eq:inelspin}
\end{align}
\end{subequations}
We find that $\sum_\alpha I_\alpha^{S_z,\mathrm{el}} \equiv 0$ vanishes at {\em all} bias voltages, so that the total spin current $\sum_\alpha I_\alpha^{S_z} = \sum_\alpha I_\alpha^{S_z,\mathrm{inel}}$ is governed by the inelastic component only which is plotted in Fig.~\ref{fig:ISV}. Thus, this quantity measures the loss of angular momentum of electrons within the interacting active region of MTJ in Fig.~\ref{fig:fig1}, which is then carried away by magnonic spin current (through the left semi-infinite lead toward the left magnonic macroscopic reservoir). Although spin current carried by electrons or magnons individually is not conserved, the total angular moment in this process is conserved. This finding further justifies the separation of currents into elastic and inelastic contributions since $\sum_\alpha I_\alpha^{S_z,\mathrm{el}} \equiv 0$ does not participate in the loss of angular momentum.

\begin{figure*}
\includegraphics[scale=0.6,angle=0]{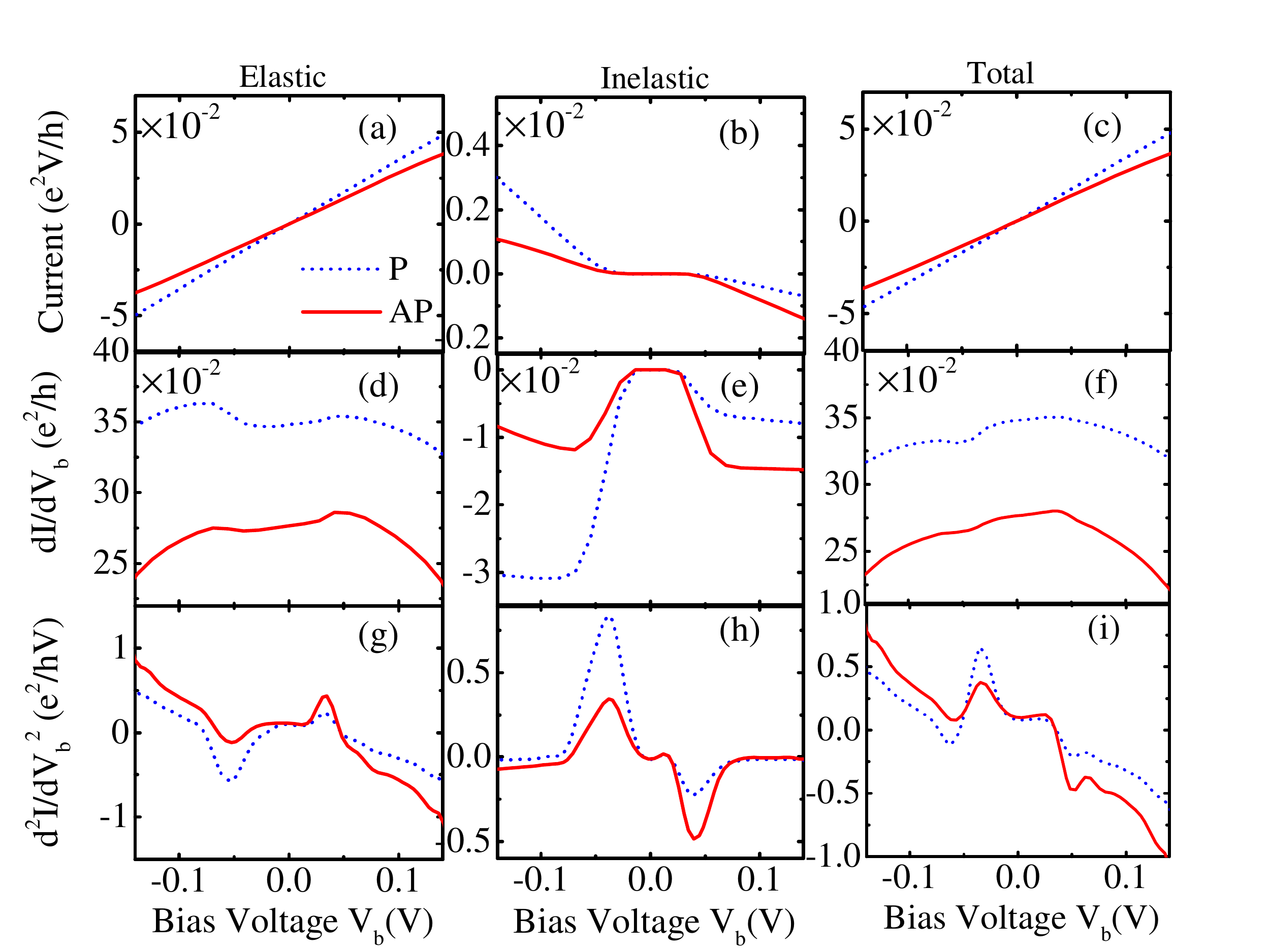}
\caption{(Color online) Elastic and inelastic charge currents in Eq.~\eqref{eq:landauer} and their sum, as well as the corresponding first and second derivatives, versus the bias voltage in the model of MTJ in Fig.~\ref{fig:fig1} with active region $N_x \times N_y \equiv 3 \times 3$  for parallel and antiparallel orientation of its magnetizations. These charge currents are obtained from the electronic GF computed in F-SCBA, which includes (in self-consistent fashion) the non-interacting magnonic GF. The temperature is set as $T=12$ K.}
\label{fig:fig8}
\end{figure*}

Due to the fact that the \mbox{e-m} interaction strength $g$ is comparable to the magnonic bandwidth, the single particle and many-body properties of magnons within the active region in Fig.~\ref{fig:fig1} are governed largely by the collective quasiparticles rather than the bare (non-interacting) magnons
we started from. This is demonstrated by plotting the magnon density of states (DOS) in Fig.~\ref{fig:fig5} within the active region versus
energy. The DOS is obtained from $-\mathrm{Tr}[\Im(\mathbf{B}^r)]/\pi$ with \mbox{e-m} interactions turned off ($g=0$) or turned on ($g \neq 0$).
In Fig.~\ref{fig:fig5}(c), we can clearly distinguish three peaks corresponding to the quasibound states suggesting the formation of long-lived
quasiparticles.  They can be interpreted as a magnon dressed by the cloud of electron-hole pair excitations out of equilibrium. Importantly for ZBA discussed below, the DOS of interacting magnons extends all the way to zero energy, thereby enabling e-m scattering even at vanishingly small bias voltage. On the other hand, the electronic DOS obtained from $-\mathrm{Tr}[\Im(\mathbf{G}^r)]/\pi$  and plotted in Figs.~\ref{fig:fig5}(a) and ~\ref{fig:fig5}(b) is only slightly perturbed when e-m interaction is turned on due to much larger electronic bandwidth.

\begin{figure}
\includegraphics[scale=0.3,angle=0]{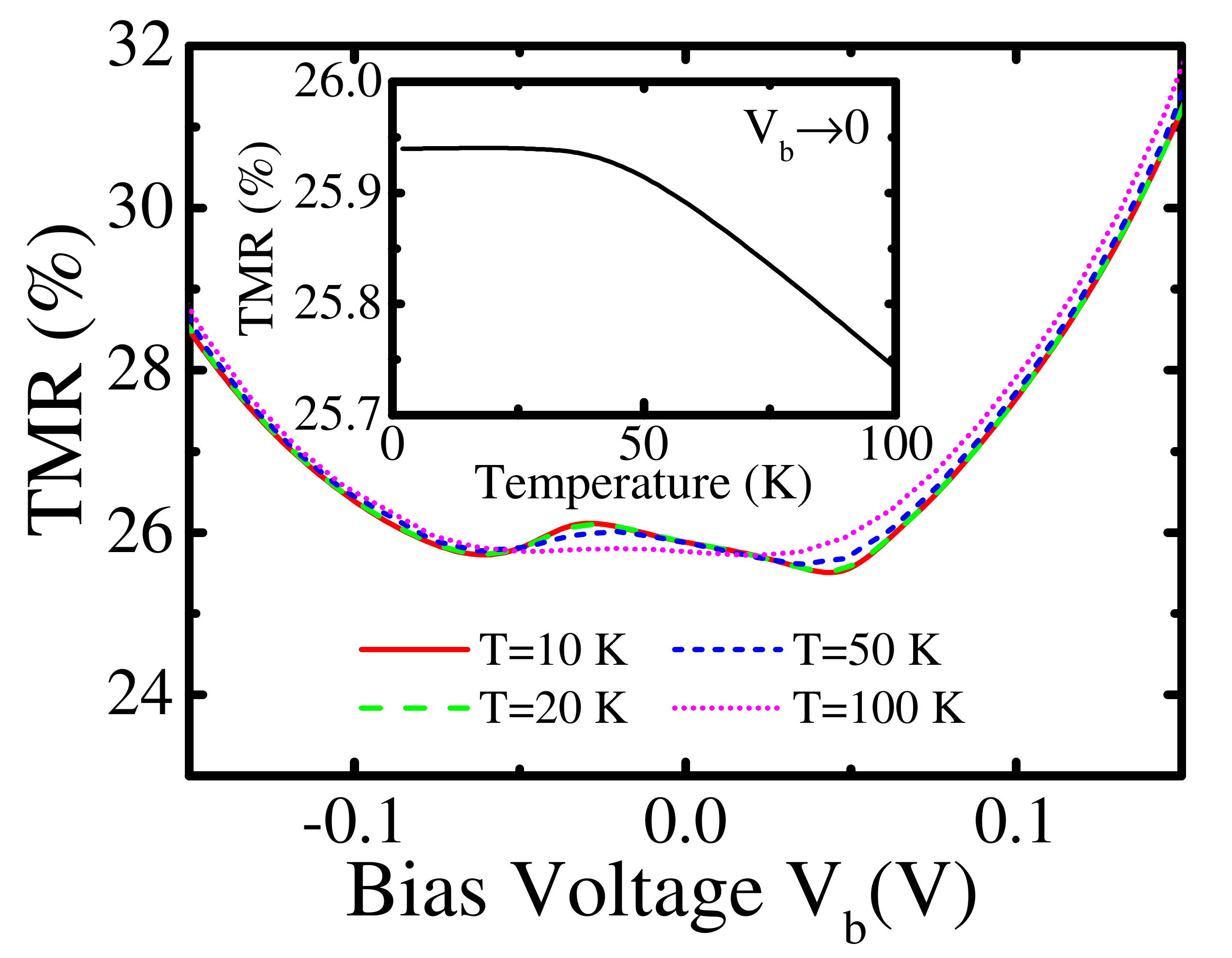}
\caption{(Color online) The TMR vs. bias voltage $V_b$ in the model of MTJ in Fig.~\ref{fig:fig1} with active region $N_x \times N_y \equiv 3 \times 3$. The inset shows TMR as a function of temperature in the linear-response limit \mbox{$V_b \rightarrow 0$}. These results are obtained from electronic GF computed by solving coupled Eqs.~\eqref{eq:eDyson}, ~\eqref{eq:mDyson} and ~\eqref{eq:fock}.}
\label{fig:fig9}
\end{figure}

Figure~\ref{fig:fig6} plots the elastic, inelastic and total charge currents, together with their first derivative $dI/dV_b$ (i.e., differential conductance) and second derivative $d^2I/dV^2_b$ (i.e., IETS~\cite{Reed2008}), as a function of the applied bias voltage $V_b$. The currents in Fig.~\ref{fig:fig6} are computed using F-SCBA for the electronic GF and self-energy of $N_x \times N_y \equiv 3 \times 1$ active region in Fig.~\ref{fig:fig1}, while the magnonic GF is used as the non-interacting one by setting $\check{\bm \Omega}_\mathrm{m-e} \equiv 0$ in Eq.~\eqref{eq:mDyson}. The inelastic current in Fig.~\ref{fig:fig6}(b) is zero until the threshold bias voltage is reached ($\simeq \pm 20$ mV according to dashed line in Fig.~\ref{fig:fig5}) at which  magnons can be excited. Above the threshold voltage, inelastic current displays Ohmic behavior. This is simply due to the fact that the rate of energy (and angular momentum) loss is proportional to the rate of electrons being injected into the active region. Although elastic current in Fig.~\ref{fig:fig6}(a) shows apparent Ohmic behavior for all bias voltages, the corresponding differential conductance in Fig.~\ref{fig:fig6}(d) deviates strongly from the straight line within the energy rage where magnons can be excited. This can be explained by the fact that the effective electronic DOS inside the active region can be changed through \mbox{e-m} scattering. The elastic differential conductance in Fig.~\ref{fig:fig6}(d) decreases once the magnons are excited, but this is compensated by the increase of inelastic differential  conductance in Fig.~\ref{fig:fig6}(e) such that the total differential conductance in Fig.~\ref{fig:fig6}(f) has less pronounced features. The IETS plotted in Figs.~\ref{fig:fig6}(g) and \ref{fig:fig6}(h) shows a clear signature~\cite{Reed2008} of the non-interacting magnonic DOS from  Fig.~\ref{fig:fig5} with two peaks emerging slightly away from $V_b=0$. Nevertheless, when these two contributions are summed up in Fig.~\ref{fig:fig6}(i), IETS for total current shows more than just two peaks.

In order to see the effect of DOS of interacting magnons (solid line in Fig.~\ref{fig:fig5}), or possible magnon heating due to tunneling electrons,  Fig.~\ref{fig:fig7} presents the same information as in Fig.~\ref{fig:fig6} but recomputed by including e-h polarization bubble diagram in Fig.~\ref{fig:fig2}(b) for the magnonic self-energy $\check{\bm \Omega}_\mathrm{m-e} \neq 0$. Since DOS of interacting magnons in Fig.~\ref{fig:fig5} is sufficiently broadened to
reach low frequencies, the inelastic current in Fig.~\ref{fig:fig7}(b) is now non-zero even for very small bias voltage $V_b \rightarrow 0$. The presence of magnons dressed by the cloud of e-h pair excitations in this calculation forces elastic conductance in Fig.~\ref{fig:fig7}(d) to increase around $V_b=0$, or inelastic conductance in Fig.~\ref{fig:fig7}(e) to decrease, which is opposite to the behavior of the same quantities in the case of non-interacting magnons analyzed in Fig.~\ref{fig:fig6}. The two peaks of opposite sign around $V_b=0$  in partial IETS plotted in Figs.~\ref{fig:fig7}(g) and ~\ref{fig:fig7}(h) look very similar to ZBA peaks observed experimentally~\cite{Drewello2008} in realistic MTJs.

Due to opposite effect of \mbox{e-m} interaction on the two partial IETS, their sum in both Figs.~\ref{fig:fig6}(i) and ~\ref{fig:fig7}(i) looses the simple two peak structure around $V_b=0$ observed experimentally.~\cite{Drewello2008} To investigate whether this complexity in the total IETS could be an artifact of 1D nature of MTJ model (with active region $N_x \times N_y \equiv 3 \times 1$ attached to 1D leads) considered in Figs.~\ref{fig:fig6} and ~\ref{fig:fig7}, we recompute the same quantities for the active region $N_x \times N_y \equiv 3 \times 3$ in Fig.~\ref{fig:fig8}. For this case, both the partial IETS in Figs.~\ref{fig:fig8}(g) and ~\ref{fig:fig8}(h) and the total IETS in Fig.~\ref{fig:fig8}(i) exhibit simple two peak structure. However, the two peaks appear  slightly away from the zero bias voltage  $V_b=0$ because we do not include (due to substantial computational expense) e-h polarization bubble diagram from Fig.~\ref{fig:fig2}(b) which is needed to introduce non-zero magnonic DOS at low energies in Fig.~\ref{fig:fig5}(d) enabling e-m scattering at
$V_b \rightarrow 0$.

The usage of $N_x \times N_y \equiv 3 \times 3$ active region introduces non-negligible $\mathrm{TMR}$ ratio which can be extracted from Fig.~\ref{fig:fig8}(c)
as $\mathrm{TMR} \simeq 26\%$ for $V_b \in (-0.1 \ \mathrm{V},0.1 \ \mathrm{V})$. Its detailed dependence on $V_b$ plotted in Fig.~\ref{fig:fig9} shows how ZBA vanishes for temperatures \mbox{$T \gtrsim 100$ K}. Also, the TMR ratio (at $V_b \rightarrow 0$) vs. temperature  shown in the inset of Fig.~\ref{fig:fig9} agrees with experimentally observed~\cite{Drewello2008,Khan2010} TMR decrease with increasing temperature.

Considering fully 3D model of MTJs, where additional $k$-point sampling is required for the transverse direction, would require carefully crafted approximations to evade prohibitively expensive five-dimensional integrals in the systems of coupled nonlinear integral equations for the electronic and magnonic GFs. Also, we note that $dI/dV_b$ in experiments~\cite{Drewello2008} has a dip at $V_b=0$, and its absolute value increases with increasing $|V_b|$ due to opening of new conducting channels by inelastic \mbox{e-m} scattering. This is not seen in Figs.~\ref{fig:fig6}(f), ~\ref{fig:fig7}(f) and ~\ref{fig:fig8}(f) due to small
number of spin-resolved conducting channels (up to six for $N_x \times N_y \equiv 3 \times 3$ active region) present in our model of MTJ.

\section{Concluding remarks}\label{sec:conclusions}

The \mbox{e-ph} interaction in nanostructures driven out of equilibrium, as the example of nonequilibrium electron-boson quantum-many body system, has been amply studied~\cite{Mitra2004,Viljas2005,Koch2006,Frederiksen2007,Galperin2007,Lee2009a,Dash2011,Urban2010,Novotny2011} over the past decade using NEGF formalism. This approach, which makes it possible to rigorously model microscopic details of inelastic scattering processes, has been typically implemented using the SCBA diagrams for the electronic self-energy, and sometimes also including e-h polarization bubble diagram for the phonon self-energy in the nonequilibrium MBPT. On the other hand, the same level of description of \mbox{e-m} scattering has received far less attention,~\cite{Reininghaus2006} despite its great relevance for a plethora of problems in spintronics.~\cite{Levy2006,Manchon2009a}

In this study, we have shown how to obtain analytical expressions for SCBA and e-h polarization bubble diagrams describing \mbox{e-m} scattering. This is achieved in a particularly compact form by using matrix GFs in the Keldysh space (which are functions of energy for electrons or frequency for magnons in steady-state nonequilibrium), thereby simplifying tracking of electron spin flips and direction of magnon propagation required to conserve angular momentum at
each vertex of the Feynman diagrams. The self-consistent solution of the corresponding system of coupled nonlinear integral equations, which is equivalent to infinite resummation of certain classes of diagrams (akin to the self-consistent GW treatment of the one-particle electronic self-energy due to electron-electron interaction out of equilibrium~\cite{Thygesen2008}), is obtained via several intertwined numerical algorithms that reduce the computational complexity of this task.

Using this framework, we have computed charge and spin currents at finite bias voltage in quasi-1D models of F/I/F MTJ illustrated in Fig.~\ref{fig:fig1}. Our {\em key results} are summarized as follows: ({\em i}) while elastic component of the sum of spin currents in all attached leads is zero at all bias voltages, the inelastic one is non-zero thereby  measuring the loss of spin angular momentum carried by magnons away from the active region (see Fig.~\ref{fig:ISV}); ({\em ii}) turning on the \mbox{e-m} interaction strongly modifies magnonic DOS, which acquires larger bandwidth while exhibiting peaks due to quasibound states of {\em magnons dressed by the cloud of electron-hole pair excitations} [see Figs.~\ref{fig:fig5}(c) and ~\ref{fig:fig5}(d)]; ({\em iii}) using F-SCBA for the electronic self-energy in Fig.~\ref{fig:fig2}(a), coupled with e-h polarization bubble diagram for the magnonic self-energy in Fig.~\ref{fig:fig2}(b), generates two peak structure around zero bias voltage in the second derivative (i.e., IETS~\cite{Reed2008}) of both elastic and inelastic charge currents (see Fig.~\ref{fig:fig7}).

We emphasize that e-h polarization bubble diagram in Fig.~\ref{fig:fig2}(b) is responsible for the substantial change of magnonic DOS (encoded by the retarded magnonic GF) in equilibrium, as well as for magnon heating (encoded by the lesser magnonic GF) in nonequilibrium due to tunneling electrons where heated magnons can also exert backaction~\cite{Novotny2011} onto electrons. Since ZBA occurs at very small bias voltages, the former effect is more important because the broadened magnonic DOS in Figs.~\ref{fig:fig5}(c) and ~\ref{fig:fig5}(d) extends to low energies (in contrast to DOS of non-interacting magnons), thereby making
possible inelastic e-m scattering even at zero bias. It is worth mentioning that the effect of electron-boson interaction on bosonic DOS has been rarely discussed in prior NEGF studies~\cite{Reininghaus2006,Mitra2004,Viljas2005,Koch2006,Frederiksen2007,Galperin2007,Lee2009a,Dash2011,Urban2010,Novotny2011} of coupled electron-boson systems, either due to simplicity of bosonic spectrum assumed or because of not inserting dressed magnonic GF lines (which include e-h polarization bubble diagram) into SCBA for electronic GF.

While partial IETS in Figs.~\ref{fig:fig7}(g) and ~\ref{fig:fig7}(h) obtained from elastic or inelastic components of charge current, respectively, are quite similar to experimentally observed~\cite{Drewello2008} ZBA, their sum in Fig.~\ref{fig:fig7}(i) is more complicated due to usage of strictly 1D model with active region $N_x \times N_y \equiv 3 \times 1$ in those Figures. Switching to $N_x \times N_y \equiv 3 \times 3$ active region attached to quasi-1D leads (supporting more than two spin-resolved conducting channels) makes total IETS in Fig.~\ref{fig:fig8}(i) exhibiting only two peaks, albeit shifted slightly away from $V_b=0$ due to exclusion of e-h polarization bubble diagram in this calculation in order to reduce computational expense.

We believe that extension of our approach to 3D MTJs [by adding computationally expensive $k$-point sampling in the $y$ and $z$ directions while keeping real space Hamiltonian from Eq.~\eqref{eq:emhamiltonian} in the $x$ direction] would be able to describe not just ZBA in realistic junctions, but also TMR and STT effects as a function of temperature and bias voltage, thereby opening a path to understand how to optimize these effects for applications in spintronics by {\em tailoring magnon spectrum}.

\begin{acknowledgments}
We thank H. Mera for immensely valuable discussions and comments on several drafts. This work was supported in part by NSF under Grant No. ECCS 1202069.
\end{acknowledgments}

\appendix

\section{Conservation of charge current in F-SCBA}\label{sec:conservation}

It is instructive to check if charge current is conserved, $\sum_{\alpha}I_{\alpha}\stackrel{?}{=}0$, after electronic GF in F-SCBA is inserted into Eq.~\eqref{eq:mw}
\begin{widetext}
\begin{align}\label{eq:proof}
\sum_{\alpha}I_{\alpha} &= \frac{e}{h} \int \!\! dE\, \mathrm{Tr} \left[ ({\bm \tau}^z \check{\bold{G}}(E) {\bm \tau}^z \check{\bold{\Sigma}}_\mathrm{leads}(E) - \check{\bold{\Sigma}}_\mathrm{leads}(E){\bm \tau}^z \check{\bold{G}}(E) {\bm \tau}^z) {\bm \tau}_{11} \right] \nonumber \\
&=-\frac{e}{h} \int \!\! dE\, \mathrm{Tr} \left[ ({\bm \tau}^z \check{\bold{G}}(E) {\bm \tau}^z \check{\bold{\Sigma}}^{F}(E) - \check{\bold{\Sigma}}^{F}(E) {\bm \tau}^z \check{\bold{G}}(E) {\bm \tau}^z) {\bm \tau}_{11} \right] \nonumber \\
&=\frac{eg^2S}{2 i h}  \int \!\! dEdE' \, \mathrm{Tr} \left[ ({\bm \tau}^z \check{\bold{G}}_{\uparrow\uparrow}(E) {\bm \tau}^z \check{\bold{B}}(E-E') \circ \check{\bold{G}}_{\downarrow\downarrow}(E')-
 \check{\bold{G}}_{\downarrow\downarrow}(E') \circ \check{\bold{B}}(E'-E) {\bm \tau}^z \check{\bold{G}}_{\uparrow\uparrow}(E) {\bm \tau}^z \right. \nonumber \\
 & \left. \ \ \ + {\bm \tau}^z \check{\bold{G}}_{\downarrow\downarrow}(E') {\bm \tau}^z \check{\bold{B}}^T(E'-E) \circ \check{\bold{G}}_{\uparrow\uparrow}(E)-
 \check{\bold{G}}_{\uparrow\uparrow}(E) \circ \check{\bold{B}}^T(E-E') {\bm \tau}^z \check{\bold{G}}_{\downarrow\downarrow}(E') {\bm \tau}^z) {\bm \tau}_{11} \right] \nonumber\\
&\equiv 0.
\end{align}
\end{widetext}
We use $\check{\bold{G}} {\bm \tau}^z (\check{\bold{\Sigma}}_\mathrm{leads} + \check{\bold{\Sigma}}^{F}) - (\check{\bold{\Sigma}}_\mathrm{leads} + \check{\bold{\Sigma}}^{F}) {\bm \tau}^z \check{\bold{G}}=0$ to write the second line in Eq.~\eqref{eq:proof}. To show that the third line is identically zero, we use the fact that three arbitrary matrices $\check{\bold{A}}$, $\check{\bold{B}}$ and $\check{\bold{C}}$ satisfy \mbox{${\tt diag}(\check{\bold{A}} \circ \check{\bold{B}} \check{\bold{C}} - \check{\bold{A}} \check{\bold{B}}^T \circ \check{\bold{C}})=0$}, where ${\tt diag}(\ldots)$ returns the main
diagonal of its matrix argument. Note that in all of the above expressions the elementwise products of the matrices are performed prior to the matrix
products. This demonstrates that no matter what approximation is employed for the magnonic GF and self-energy, the charge current will be conserved as long as the self-consistency is achieved when computing electronic GF and self-energy within F-SCBA.

\section{Numerical implementation of the Hilbert transform}\label{sec:ap_hilbert}

The Hilbert transform of a function $f(x)$ is defined by
\begin{equation}\label{eq:hilbert}
\mathcal{H}[f(y)] = \frac{1}{\pi} \mathcal{P} \int\limits_{-\infty}^{\infty} dx \frac{f(x)}{y-x}.
\end{equation}
Due to nonanalytic nature of the integrand, performing numerical integration directly over a mesh of equidistant points, or even over an adaptive mesh, is very time consuming. Instead, other approaches, typically based on the fast Fourier transform (FFT), are used to speed up~\cite{Frederiksen2007,Dash2011} the computation. However, using FFT requires $f(x)$ to be defined on the mesh of equidistant points. In our calculations $f(x)$ is given on the mesh of adaptively selected points with second order polynomial interpolation. Therefore, here we explain an algorithm that can give Hilbert transform for this type of input function.

Let us assume that the function $f(x)$ is given on $2N+1$ adaptively chosen points $x_n$, such that the value of the function inside the
range $(x_{2n-1},x_{2n+1})$ is obtained by a second order polynomial interpolation using three points $(x_{2n-1},x_{2n},x_{2n+1})$.
In this case, the integral in Eq.~\eqref{eq:hilbert} can be rewritten as
\begin{widetext}
\begin{align}
\mathcal{H}[f(x)]=\frac{1}{\pi}\sum_{n=1}^N \mathcal{P} \int\limits_{x_{2n-1}}^{x_{2n+1}} dy \frac{a_{n}(y)f_{2n-1}+b_{n}(y)f_{2n}+c_{n}(y)f_{2n+1}}{x-y},
\end{align}
where
\begin{subequations}
\begin{align}
&a_n(y)=\frac{(y-x_{2n})(y-x_{2n+1})}{(x_{2n-1}-x_{2n})(x_{2n-1}-x_{2n+1})},\\
&b_n(y)=\frac{(y-x_{2n-1})(y-x_{2n+1})}{(x_{2n}-x_{2n-1})(x_{2n}-x_{2n+1})},\\
&c_n(y)=\frac{(y-x_{2n-1})(y-x_{2n})}{(x_{2n+1}-x_{2n-1})(x_{2n+1}-x_{2n})}.
\end{align}
\end{subequations}
The integration can then be performed analytically to obtain
\begin{align}
\mathcal{H}[f(y)]=\frac{1}{\pi}\sum_{n=1}^N f_{2n-1}a^h_{n}(y)+f_{2n}b^h_{n}(y)+f_{2n+1}c^h_{n}(y),
\end{align}
where
\begin{subequations}\label{eq:arrays}
\begin{align}
& a^h_{n}(y) = \frac{\frac{1}{2}(x_{2n+1}-x_{2n-1})-(y-x_{2n})-\frac{(y-x_{2n})(y-x_{2n+1})}{x_{2n+1}-x_{2n-1}}L_n}{x_{2n}-x_{2n-1}}, \\
& b^h_{n}(y) = \frac{(y-x_{2n+1})^2-(y-x_{2n-1})^2-2(y-x_{2n-1})(y-x_{2n+1})L_n}{2(x_{2n}-x_{2n-1})(x_{2n}-x_{2n+1})}, \\
& c^h_{n}(y) = \frac{-\frac{1}{2}(x_{2n+1}-x_{2n-1})-(y-x_{2n})-\frac{(y-x_{2n-1})(y-x_{2n})}{x_{2n+1}-x_{2n-1}}L_n}{x_{2n+1}-x_{2n}}, \\
& L_n = \log \left( \frac{y-x_{2n+1}}{y-x_{2n-1}} \right).
\end{align}
\end{subequations}
The arrays in Eq.~\eqref{eq:arrays} can be combined into a single array $h_n(y)$
\begin{align}
&h_1(y)=a_1^h(y), \ \ \ h_{2N+1}(y)=c_N^h(y), \nonumber \\
&h_{2n}(y)=b_{n}^h(y), \ \ \ h_{2n+1}(y)=a_{n+1}^h(y)+c_n^h(y),
\end{align}
so that Hilbert transform is computed using the following formula
\begin{align}
\mathcal{H}[f(y)]=\frac{1}{\pi} \sum_{n=1}^{2N+1} f_n h_n(y).
\end{align}
\end{widetext}

%BibTeX
%Windows:
%\bibliographystyle{D:/PHYSICS/TEX/BIBTEX/prsty}
%\bibliography{D:/PHYSICS/TEX/BIBTEX/qttg}

%Linux:
%\bibliographystyle{apsrev}
%\bibliography{qttg}

\end{document}